\documentclass[aps,pre,showpacs,twocolumn,preprintnumbers,amsmath,amsfonts,bm]{revtex4}
\usepackage[dvips]{graphicx}
\usepackage{dcolumn}

\begin{document}
\title{Critical Line in Random Threshold Networks with Inhomogeneous Thresholds}

\author{Thimo Rohlf}

\affiliation{Santa Fe Institute, 1399 Hyde Park Road, Santa Fe, NM 87501, U.S.A }
\date{\today}

\begin{abstract}
We calculate analytically the critical connectivity $K_c$ of Random Threshold Networks (RTN) 
for homogeneous and
inhomogeneous thresholds, and confirm the results by numerical simulations. We
 find a super-linear increase of $K_c$ with the (average) absolute threshold $|h|$, which approaches $K_c(|h|) \sim h^2/(2\ln{|h|})$ for large $|h|$,
and show that this asymptotic scaling is universal for RTN with Poissonian distributed connectivity and
threshold distributions with a variance that grows slower than $h^2$. Interestingly, we find that 
inhomogeneous distribution of thresholds leads to increased propagation of perturbations
for sparsely connected networks, while for densely connected networks damage
is reduced; the cross-over point yields a novel, characteristic connectivity $K_d$, that has no
counterpart in Boolean networks.
Last, local correlations between node thresholds and in-degree are introduced. Here, numerical simulations show
that even weak (anti-)correlations can lead to a transition from ordered to chaotic dynamics, and vice versa.
It is shown that the naive mean-field assumption typical for the annealed approximation leads to false predictions
in this case, since correlations between thresholds and out-degree that emerge as a side-effect strongly modify
damage propagation behavior.
 
\end{abstract}


\maketitle

\section{Introduction}
Many systems in nature, technology and society can be described as
complex networks with some flow of matter, energy or information between
the entities the system is composed of; examples are neural networks,
gene regulatory networks, food webs, power grids and friendship networks.
Often, in particular when the networks considered are very large, many details
of the topological structure as well as of the dynamical interactions between
units are unknown, hence, statistical methods have to be applied
to gain insight into the global properties of these systems.
In this spirit, Kauffman \cite{Kauffman69,Kauffman93} introduced
the notion of Random Boolean Networks (RBN), originally as a simplified model
of gene regulatory networks (GRN). In a RBN of size $N$, each node $i$ receives inputs from $0 \le k \le N$
other nodes (with $k$ usually either considered to be constant, or distributed
according to a Poissonian with average $\bar{K} \ll N$), and updates its
state according to a Boolean function $f_i$ of its inputs; the subscript $i$ indicates
that Boolean functions vary from site to site, usually assigned at random to each
node. It was shown that RBN exhibit a percolation transition from ordered to chaotic
dynamics at a critical connectivity $\bar{K} = K_c = 2$.
Since interactions in RBN are asymmetric and hence a Hamiltonian does not exist,
mean-field techniques have to be applied for analytical calculation
of critical points,
for example the so-called {\em annealed approximation (annealed approximation)} introduced by Derrida and Pomeau \cite{DerridaP86,SoleLuque95,LuqueSole96}.
In the annealed approximation, random perturbations are applied to initial dynamical states, and random ensemble
techniques are applied to determine whether the so-induced "damage" spreads over the network or not. 
Recent research has revealed many surprising details of RBN dynamics at criticality, e.g.
super-polynomial scaling of the number of different dynamical attractors (fixed points
or periodic cycles) with $N$ \cite{SamuelsonTroein03} (while Kauffman assumed it to scale $\sim \sqrt{N}$ \cite{Kauffman69}),
as well as analytically derived scaling laws for mean attractor periods \cite{AlbertBaraBoolper00} and for the
number of frozen and relevant nodes in RBN \cite{KaufmanMihaljevDrossel05,Mihaljev2006}. Similarly, it was shown recently that
dynamics in finite RBN exhibits considerable deviations from the annealed approximation (that is exact only in the
limit $N\to\infty$) \cite{RohlfGulbahceTeuscher2007,leone2006}. Boolean network models have been applied successfully to model
the dynamics of real biological systems, e.g. the segment polarity network of {\em Drosophila} \cite{AlbertOthmer2003},
dynamics and robustness of the yeast cell cycle network \cite{Braunewell2007}, 
damage spreading in knock-out experiments \cite{RamoeKesseliYli06} as well as establishment of position
information \cite{RohlfBornhol2005} and cell differentiation \cite{Jackson1986} in development. Other models
explicitly evolve RBN topology according to local rewiring rules coupled to local order parameters
of network dynamics (e.g., the local rate of state changes), and investigate the
resulting self-organized critical state \cite{BornholRohlf00,BornholRoehl2003,LiuBassler2006}.

A drawback of RBN is the fact that, in spite of their discrete nature (which makes them easy to simulate
on the computer in principle), the time needed to compute their dynamics in many instances scales
exponential in $N$ and $\bar{K}$, and often large statistical ensembles are needed 
for unbiased statistics due to the strongly non-ergodic
character \cite{Moreira2005} of RBN dynamics. For this reason, there exists considerable interest in simplified
models of RBN dynamics, as, for example, {\em Random Threshold Networks} (RTN), that constitute
a subset of RBN.

In RTN, states of network nodes are updated according to a weighted sum of their inputs plus a threshold $h$, while
interaction weights take (often discrete and binary) positive or negative values assigned at random.  
The critical connectivity, calculated by means of the annealed approximation, was found to deviate slightly from RBN
\cite{Kuerten88a,Kuerten88b,RohlfBornhol02}; this analysis was extended to RTN dynamics including stochastic update errors \cite{Nakamura2004}.
In particular, it was found that phase transitions in  RTN with scale-free topologies \cite{Nakamura2004,Aldana2004}  substantially differ
from both RTN with homogeneous or Poissonian distributed connectivity and scale-free RBN \cite{Aldana2003}. Further,
dynamics in finite RTN with $k = const. =2$ inputs per node recently was found to be surprisingly ordered,
including, e.g., globally synchronized oscillations \cite{Greil2007}. Other approaches, that apply
learning algorithms as well as ensemble techniques, present evidence that information processing
of static \cite{Patarnello1989} or time-variant \cite{Bertschi2004} external inputs is optimized at criticality in both RBN and RTN.

In this paper, we extend the theoretical analysis of RTN in  a number of respects. First, we calculate the
critical connectivity $K_c$ for arbitrary thresholds $h \le 0$, and generalize this derivation for the first time to
inhomogeneously distributed thresholds $h_i$ that can vary from node to node. This generalization, that introduces
an additional level of complexity to RTN dynamics, is motivated by recent observations of strong variations in regulatory
dynamics from gene to gene in real GRN, caused by, for example, the frequent occurrence of canalizing functions \cite{Moreira2005}
and the abundance of regulatory RNA in multicellular organisms which strongly influence the expression levels and -patterns of
(regulatory) proteins \cite{StadlerEvoRNA2005}. Using the annealed approximation and additional approximation techniques, we derive a general
scaling relationship between critical connectivity $K_c$ and (average) absolute node threshold $|h|$, and show
that $K_c(|h|)$ asymptotically approaches a unique scaling law $K_c(|h|) \sim h^2/(2\ln{|h|})$ 
for large $|h|$. Evidence is presented that this asymptotic scaling law is universal for RTN with Poissonian distributed
connectivity and threshold distributions with a variance that grows slower than $|h|^2$. Convergence
against this scaling law is rather slow (logarithmic in $|h|$); we show that, for finite $|h|$, scaling behavior
can be approximated well locally by power laws $K_c(|h|) \sim |h|^{\alpha}$ with $3/2 < \alpha < 2$.

Further, we establish that damage propagation functions of RTN with homogeneous thresholds $|h|$ and of RTN with inhomogeneous thresholds
with the same {\em average} $\bar{|h|} = |h|$ intersect at characteristic connectivities $K_d(|h|) > K_c(|h|)$, which implies that for $\bar{K} < K_d$, random
distribution of thresholds tends to increase damage, while for $\bar{K} > K_d$, the opposite holds. Evidence is
presented that $K_d(|h|)$ converges to an asymptotic scaling law  $K_d(|h|) \sim h^2$ . We compare
the scaling of $K_d$ to the corresponding case of random Boolean networks (RBN) with inhomogeneously distributed bias,
parameterized in terms of a bias parameter $1/2 \le p \le 1$. It is shown that
$K_d$ {\em is not defined for RBN} in the limit $p \to 1$, which
corresponds to $|h| \to \infty$ in RTN. Hence, $K_d$ constitutes a truly novel, not previously known concept, yielding a new characteristic connectivity which is well-defined only for RTN.

Last, we investigate the effect of correlations between thresholds $h_i$ and in-degree $k_i$, while keeping all
other network parameters constant. We find that even small positive correlations can induce a transition from
supercritical (chaotic) to subcritical (ordered) dynamics, while anti-correlations have the opposite effect.
It is shown that the naive mean-field assumption typical for the annealed approximation leads to false predictions
in this case. Even in the simplest case, where only in-degree and (absolute) threshold are correlated,
complete information about topology, including the output side, has to enter
statistics, and the order of averages becomes important.

%\vfill
\section{Random Threshold Networks}
A Random Threshold Network (RTN) consists of $N$ randomly interconnected binary 
sites (spins) with states $\sigma_i=\pm1$.
For each site $i$, its state 
at time $t+1$ is a function of the inputs it receives from other 
spins at time $t$:
\begin{eqnarray} 
\sigma_i(t+1) = \mbox{sgn}\left(f_i(t)\right) \label{sign_eqn}
\end{eqnarray}  
with 
\begin{eqnarray} 
f_i(t) = \sum_{j=1}^N c_{ij}\sigma_j(t) + h_i,  \label{f_i_eqn}
\end{eqnarray}
where $c_{ij}$ are the interaction weights. 
If $i$ does not receive signals from $j$, one has $c_{ij} = 0$,
otherwise, interaction weights take discrete values $c_{ij}=\pm 1$, $+1$ or $-1$ with equal probability. 
In the following discussion we assume that 
the threshold parameter takes integer values $h_i \le 0$ 
\footnote{We restrict ourselves to negative (or zero) thresholds, to ensure
that the 'default state' of a network site $i$, i.e. when its inputs sum to zero, is to be 'inactive' ($\sigma_i = -1$), which naturally excludes positive thresholds.}.
Further, we define $\mbox{sgn}(0) = -1$. 
\footnote{Other authors define  $\mbox{sgn}(0) = +1$, however, for symmetry reasons update dynamics is not affected by
either choice. If we interpret the state $\sigma_i = -1$ as 'inactive' and, correspondingly, $+1$ as 'active', our choice appears to be more natural:
the default state of a network site is to be 'inactive', unless it receives activating inputs from other sites.}
The $N$ network sites are updated \emph{synchronously}.
Notice that we depart from the well-studied case $h_i = const. = 0$ in two
respects: $h_i$ can take arbitrary values $h_i \le 0$, and it can differ from
node to node (inhomogeneous thresholds).\\

Let us now have a closer look on network topology. 
Let $\bar{K}$ be the {\em average connectivity}, i.e. the average number of inputs (outputs) per site, and let us assume 
that each interaction weight has equal probability $p = \bar{K}/N$ to take a non-zero value. Further, let us
consider the limit of sparsely connected networks with $\bar{K} \ll N$.
Under these assumptions, the statistical distribution $\rho_k$ of in- and out-degrees follows a Poissonian: 

 \begin{equation} \rho_k = \frac{\bar{K}^k}{k!}\,e^{-\bar{K}}. \end{equation}

Further, we study the case where in- and out-degree distributions differ: while the out-degree is still distributed according to a Poissonian,
the in-degree distribution exhibits a power-law tail, i.e.
\begin{equation}
\rho_{k_{in}} \propto k^{-\gamma} 
\end{equation}
with $2 \le \gamma \le 4$.

\section{Calculating the critical line}
\subsection{Uniform threshold $h < 0$}
We start with the simplest case and assume that all network sites have identical integer threshold
values $h_i \equiv h  \le 0$. The case $h > 0$ is not studied here, as it may lead to the pathological outcome
of nodes set to an active state $\sigma_i = +1$, though they receive only inhibitory inputs $c_{ij} < 0$.

\begin{figure}[htb]
\begin{center}
\resizebox{85mm}{!}{\includegraphics{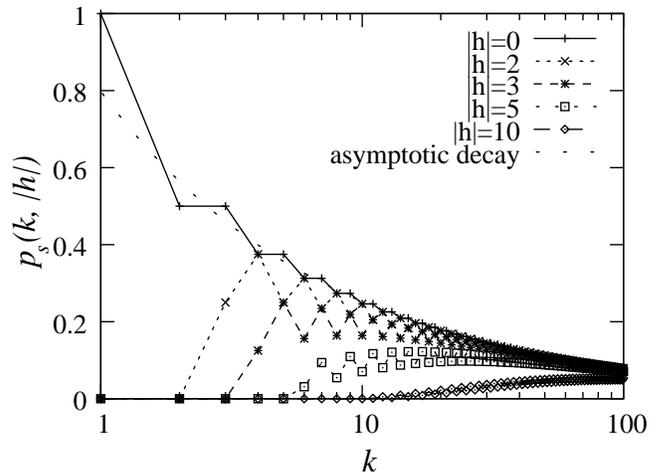}}
\end{center}
\caption{\small Probability $p_s(k, |h|)$ of damage propagation, for different values of the threshold $|h|$, as a function
of the number of inputs $k$. For large $k$, the curves asymptotically approach $p_s \sim 1/\sqrt{k}$ (dashed line). Notice the oscillatory behavior
for $|h| > 0$.}
\label{psofh_fig} 
\end{figure}

Let us first calculate the probability for damage spreading $p_s(k)$, i.e. the probability that a
node with $k$ inputs changes its state, if one of its input states is flipped. A straight-forward
extension of the combinatorial analysis carried out in \cite{RohlfBornhol02} for the special case $h = 0$ yields
\begin{eqnarray}
p_s(k,|h|) &=& k^{-1}\cdot 2^{-(k+1)}\cdot\left[ (k+|h|+1)\cdot \binom{k}{\frac{k+|h|+1}{2}}\right. \nonumber\\
& &\left. + (k-|h|+1)\cdot\binom{k}{\frac{k-|h|+1}{2}}\right] \\
&=& 2^{-(k-1)}\binom{(k-1)}{\frac{k+|h|-1}{2}} \label{odd_ps_eqn}
\end{eqnarray}
for odd $k-|h|$ with $k > |h|$, and
\begin{eqnarray}
p_s(k,|h|) &=& k^{-1}\cdot 2^{-(k+1)}\cdot\left[ (k-|h|)\cdot \binom{k}{\frac{k-|h|}{2}}\right. \nonumber\\
& &\left. + (k+|h|+2)\cdot\binom{k}{\frac{k+|h|+2}{2}}\right] \\
&=& 2^{-(k-1)}\binom{(k-1)}{\frac{k+|h|}{2}} \label{even_ps_eqn}
\end{eqnarray}
for even $k-|h|$ with $k > |h|$ (for a detailed derivation, please refer to appendix A).
Notice that Eqs. (\ref{odd_ps_eqn}) and (\ref{even_ps_eqn}) are similar, yet not identical to the corresponding relations derived in \cite{Nakamura2004}
for RTN with probabilistic time evolution; in particular, for the RTN with {\em deterministic} dynamics as studied here, the relation $p_s^{odd}(k) = p_s(k-1)$
holds only for the special case $|h| = 0$, whereas for $|h| > 0$, $p_s(k)$ exhibits an oscillatory behavior (Fig. \ref{psofh_fig}).

\begin{figure}[htb]
\begin{center}
\resizebox{85mm}{!}{\includegraphics{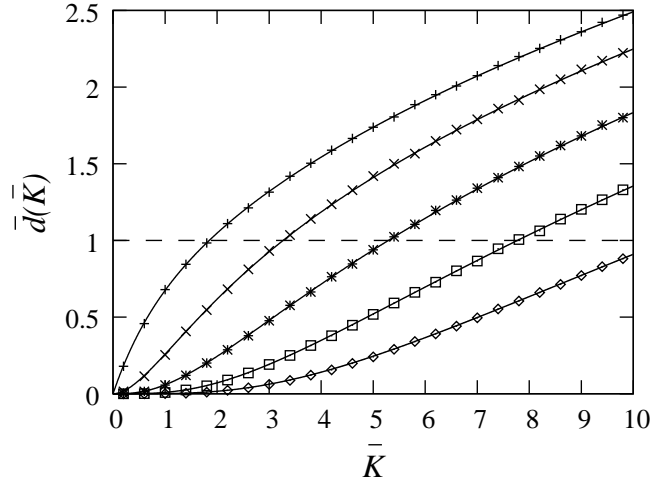}}
\end{center}
\caption{\small Expectation value $\bar{d}$ of damage one time step after a one-bit perturbation, as a function of the average connectivity $\bar{K}$,
and different (homogeneous) thresholds $|h|$ ($|h|=0$ (+), $|h|=1$ (X), $|h|=2$ (*), $|h|=3$ ($\Box$), $|h|=4$ ($\Diamond$).
Solid curves are the corresponding analytical results obtained from the annealed approximation.}
\label{averdam_fig} 
\end{figure}

\begin{figure}[htb]
\begin{center}
\resizebox{85mm}{!}{\includegraphics{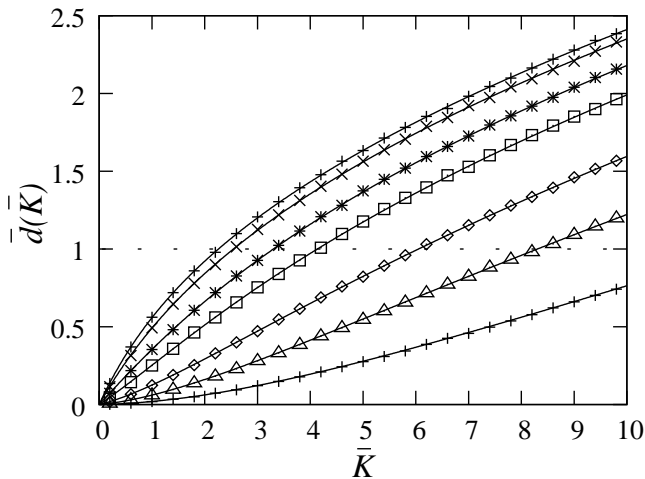}}
\end{center}
\caption{\small Average damage $\bar{d}(\bar{K})$ one time step after a one-bit perturbation, for Poisson-distributed
connectivity with average degree $\bar{K}$, and Poisson-distributed negative thresholds with average absolute value
$\bar{|h|}$; points are data from numerical simulations of RTN (ensemble averages over 100000 different network 
realizations for each data point), lined curves are analytical solutions (annealed approximation). Numerical data where
sampled for $\bar{|h|} = 0$ (+), $\bar{|h|} = 0.3$ (X), $\bar{|h|} = 1.0$ (*), $\bar{|h|} = 1.5$ (squares), 
$\bar{|h|} = 2.5$ ($\diamond$), $\bar{|h|} = 3.5$ (triangle) and $\bar{|h|} = 5.0$ (+).}
\label{averdam_inhom_fig} 
\end{figure}

\begin{figure}[htb]
\begin{center}
\resizebox{85mm}{!}{\includegraphics{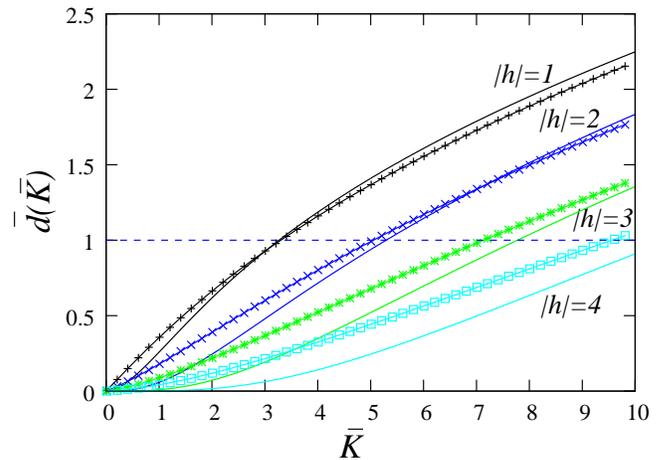}}
\end{center}
\caption{\small Comparison of damage spreading in networks with homogenenous thresholds $|h| =const.$ (solid lines, threshold values
$|h|$ as indicated) vs.
networks with inhomogeneous thresholds distributed according to a Poissonian with the same {\em average} threshold $\bar{|h|}$ (curves
with data points, $\bar{|h|} = 1$ (+), $\bar{|h|} = 2$ (x), $\bar{|h|} = 3$ (*) and  $\bar{|h|} = 4$ ($\square$); results obtained from the annealed approximation.  }
\label{dam_ihhcomp_fig} 
\end{figure}

If we know the statistical distribution function $\rho_k$ of the in-degree, the average
damage spreading probability then simply follows as \cite{RohlfBornhol02}
\begin{equation}
\langle p_s\rangle = \sum_{k=|h|}^N \rho_k \,p_s(k+1,|h|) \label{general_spread_eqn},
\end{equation}
where $\langle . \rangle$ indicates the average over the ensemble of all possible network topologies that can be generated
according to the degree-distribution  $\rho_k$.
In the case of a Poisson distributed connectivity with average degree degree $\bar{K}$, 
it follows
\begin{equation}
\langle p_s\rangle(\bar{K},|h|) = e^{-\bar{K}}\sum_{k=|h|}^N \frac{\bar{K}^k}{k!} \,p_s(k+1,|h|). \label{ER_spread_eqn}
\end{equation}
Let us now apply the so-called {\em annealed approximation} \cite{DerridaP86}, which averages the effect of perturbations
over the whole ensemble of possible network topologies {\em and} all possible state configurations;
in this approximation, the expected damage $\bar{d}$ after one update time step, given a
one-bit perturbation at time $t-1$ then follows as
\begin{equation}
\bar{d}(t+1) = \langle p_s\rangle(\bar{K},|h|)\cdot\bar{K}, \label{averdam_exp_eqn}
\end{equation}
where $\bar{.}$ denotes the average over all possible network topologies and all possible state configurations.
If we apply a sufficiently large (but finite) upper limit $N$ to the sum in Eq. (\ref{ER_spread_eqn}), we can numerically evaluate this formula
with any desired accuracy.
 Figure \ref{averdam_fig} shows
the results for the first five values of negative $h$ of RTN with Poissonian distributed
connectivity, compared to measurements obtained from numerical simulations of large ensembles
of randomly generated instances of RTN, indicating an excellent match between theory and simulation.

\subsection{Poisson distributed thresholds}
Let us now consider the more general case of non-uniform thresholds, i.e., networks where each site $i$ has
assigned an individual threshold $h_i \le 0$. In the simplest case, we can imagine that the final thresholds
resulted from iterated, random decrementations (starting from $h=0$ for all sites), until a certain average
threshold $\bar{h}$ is reached - this process results in Poisson distributed thresholds $h_i$. If threshold
assignment is independent from the (also Poisson distributed) in-degree, the probabilities for $k$ and $h$
simply multiply, and the resulting average damage propagation probability is
\begin{equation}
\langle p_s\rangle(\bar{K},\bar{|h|}) 
= e^{-(\bar{K}+\bar{|h|})}\sum_{|h|=0}^N\sum_{k=|h|}^N \frac{\bar{K}^k\bar{|h|}^{|h|}}{k!|h|!} \,p_s(k+1,|h|), \label{damspread_poih_eqn}
\end{equation}
where $\bar{|h|}$ is the average absolute threshold.

Figure \ref{averdam_inhom_fig} demonstrates that the expected damage $\bar{d}_{t+1}(\bar{K}, \bar{|h|})$ resulting from a one-bit perturbation 
at time $t$, as predicted from this annealed approximation over both degree- and threshold distribution, exhibits excellent agreement with
the results obtained from numerical simulations of randomly generated RTN ensembles. 
It is an interesting question how the dynamics of RTN with inhomogeneous thresholds compares to RTN with homogeneous 
thresholds. Figure \ref{dam_ihhcomp_fig} shows $\bar{d}(\bar{K})$ for RTN with different homogeneous $|h| =const.$ and the corresponding inhomogeneous
RTN with Poisson-distributed thresholds with the same {\em average } $\bar{|h|} = |h|$, as obtained from the annealed approximation.
One observes that for small $\bar{K}$, the curves for RTN with inhomogeneously distributed thresholds are systematically above those
of the corresponding homogeneous RTN, i.e., the randomization of node thresholds increases dynamical disorder - also, the critical
connectivities $K_c(|h|)$ (intersections with the line $\bar{d} = 1$) are shifted to smaller values. However, one also realizes
that the curves intersect in the supercritical phase at characteristic connectivities $K_d(|h|)$, i.e., for $\bar{K}> K_d(|h|)$,
inhomogeneity in thresholds actually {\em reduces} damage. 

\begin{figure}[htb]
\begin{center}
\resizebox{85mm}{!}{\includegraphics{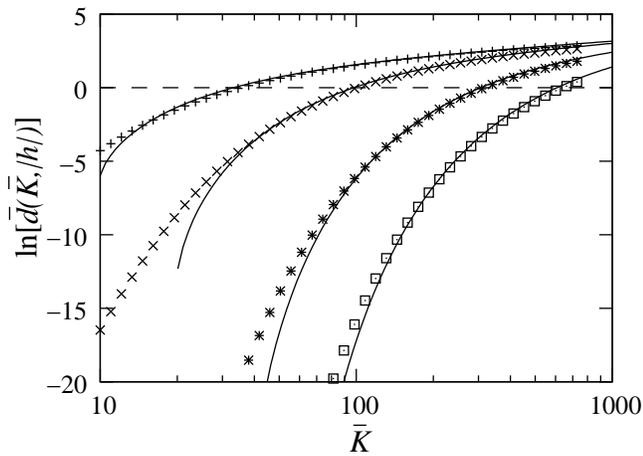}}
\end{center}
\caption{\small Logarithm of the average damage, $\ln{[\bar{d}(\bar{K})]}$, as calculated from the annealed approximation, for
different values of $|h|$ ($|h|=10$ (+), $|h| = 20$ (X), $|h| = 40$ (*) and $|h| = 60$ ($\Box$)). The corresponding solid curves are obtained from Eq. 
(\ref{log_approx_eqn}). For not to small $\bar{K}$,
one finds that Eq. (\ref{log_approx_eqn}) approximates the true damage function very well.}
\label{dcomp_fig} 
\end{figure}

\begin{figure}[htb]
\begin{center}
\resizebox{85mm}{!}{\includegraphics{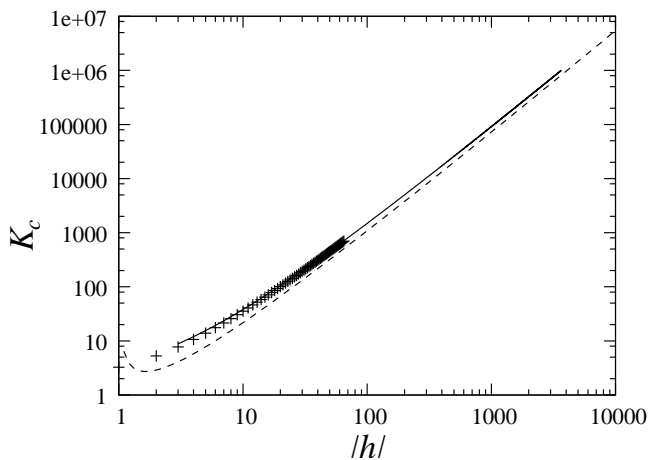}}
\end{center}
\caption{\small Scaling behavior of the critical connectivity $K_c(|h|)$ as a function of the (homogeneous)
node threshold $|h|$, log-log-plot. Data points $+$ are solutions obtained from the annealed approximation of Eq. (\ref{critdef_eqn}), the solid
curve is obtained from setting Eq. (\ref{log_approx_eqn}) to zero. The dashed line shows the asymptotic scaling behavior stated in Eq. ().}
\label{kcofh_fig} 
\end{figure}

\subsection{Universal scaling of the critical line}
\begin{figure}[htb]
\begin{center}
\resizebox{85mm}{!}{\includegraphics{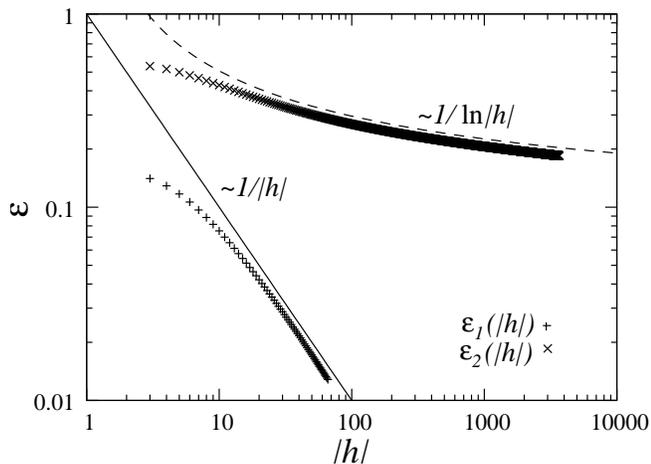}}
\end{center}
\caption{\small Crosses (X): Relative error $\varepsilon_1$  between the
approximation of Eq. (\ref{log_approx_eqn}) and the result obtained from the annealed approximation, as a function of $|h|$. For  $|h| \ge 15$,
$\varepsilon_1$ vanishes $\propto |h|^{-1}$; straight line with slope $-1$ shown for comparison. 
Data points (+): Relative error $\varepsilon_2$
 between the approximation of Eq. (\ref{log_approx_eqn}) and the asymptotic scaling of Eq. (\ref{asymp_eqn}); 
$\varepsilon_2$ goes to zero logarithmically (compare to dashed curve).}
\label{kerror_fig} 
\end{figure}

If we again assume a one-bit perturbation at time $t$,
the critical line $K_c(|h|)$, that separates the ordered and the chaotic phase of RTN dynamics, is given by the
condition 
\begin{equation}
\bar{d}(t+1) = \langle p_s\rangle(K_c(|h|),|h|)\cdot K_c(|h|) = 1. \label{critdef_eqn}
\end{equation}
Again, we can apply Eq. (\ref{ER_spread_eqn}) to solve this equation for arbitrary $h \le 0$, however, numerical evaluation
is almost impossible for $|h| > 80$ due to exponentially diverging computing time caused by evaluation of the
sum in Eq.  (\ref{critdef_eqn}) for large $\bar{K}$ \footnote{To obtain accurate results,
one has to consider networks sizes $N \gg \bar{K}$, and adjust the upper limit of the sum in (\ref{ER_spread_eqn})
accordingly. Since a small step size $\Delta\bar{K}$ has to be applied iteratedly to identify $K_c$, this becomes
computationally very costly.}. For estimation of the scaling behavior of $K_c(|h|)$ for larger $|h|$,
we are interested in a good approximation that does not require summation over the whole network topology,
and hence neglect the variation in $k$, considering damage propagation in the {\em mean field limit}
 $k = const. \approx \bar{K}$ (for details, see
Appendix B). Using the Stirling approximation: 
$n! \approx n^n e^{-n} \sqrt{2\pi n}$,
this leads
to the following approximation for the logarithm of the damage:
\begin{eqnarray}
\ln{[\bar{d}(\bar{K}, |h|)]} \approx && \left.\frac{1}{2}\,\right\{\ln{\bar{K}}%\nonumber\\
-\bar{K}\cdot\ln{\left[1-\left(\frac{|h|}{\bar{K}}\right)^2  \right]} \nonumber \\
&-&\left.|h|\ln{\left[\frac{\bar{K}+|h|}{\bar{K}-|h|}\right]}\right\} + C \label{log_approx_eqn}
\end{eqnarray}
with $C = \ln{\left(\sqrt{2/\pi}\right)}$; solving this equation for 
\begin{equation}
\ln{[\bar{d}(K_c(|h|), |h|]} = 0 \label{log_crit_eqn}
\end{equation}
then yields the critical connectivity
$K_c(|h|)$. Figure \ref{dcomp_fig} shows that this approximation is very accurate even for considerably small, finite $|h|$.
In particular, one can show that for  $|h| \ge 10$ the relative error $\epsilon$ between the
approximation of Eq. (\ref{log_crit_eqn}) and the result obtained from the annealed approximation 
vanishes $\sim |h|^{-1}$ (Fig. \ref{kerror_fig} ).

%%%%%%%%%%%new: discussion of asymptotic scaling

Still, Eq. (\ref{log_crit_eqn}) has to be solved numerically to calculate $K_c(|h|)$, and hence does not
yield information about the scaling behavior in the limit $|h| \to \infty$.
A first insight into the expected scaling can be obtained from an analysis of the scaling behavior
of the maximum of $p_s(k, |h|)$ with respect to $|h|$; if we restrict our analysis to even $k-|h|$, $k_{max}$ is given by the
condition
\begin{equation}
\Delta p_s = p_s(k, |h|) - p_s(k-2,|h|), \approx 0 \label{max_cond_eqn}
\end{equation}
or, more accurately, we have to find the minimum of the absolute value $|\Delta p_s/\Delta k|$ of the 'discrete derivative' of
$p_s(k, |h|)$ for even $k-|h|$, with $\Delta k = const. = 2$.
Inserting Eq. (\ref{even_ps_eqn}) then yields
\begin{eqnarray}
\Delta p_s &=& 2^{-k+3}\frac{(k-3)!}{[(k+|h|-3)/2]![(k-|h|-3)/2]!}\cdot  \\ \nonumber
          &&\cdot \left\{ \frac{(k-1)(k-2)}{(k+|h|+1)(k-|h|-1) }     -1          \right\}.
\end{eqnarray}
Obviously, the pre-factor on the right hand-side is always positive;
consequently, in order to determine the maximum of $p_s(k, |h|)$, we have to solve the  equation
\begin{equation}
\frac{(k-1)(k-2)}{(k+|h|+1)(k-|h|-1) } -1  = 0.
\end{equation}
Using simple algebra, one can show that
\begin{equation}
k_{max} = |h|^2 +1
\end{equation}
solves this equation, i.e. the maximum of $p_s(k, |h|)$ scales quadratically with $|h|$. 
Since $p_s(k, |h|)$ for $|h| \gg 0$ vanishes both for small and large $k$, it is plausible
that the scaling behavior of $K_c$ is dominated by the leading behavior of the maximum of the
distribution, i.e. should scale $\sim f(|h|)|h|^2$, where 
contributions from the tails of the distribution are considered in $f(|h|)$.

\begin{figure}[htb]
\begin{center}
\resizebox{85mm}{!}{\includegraphics{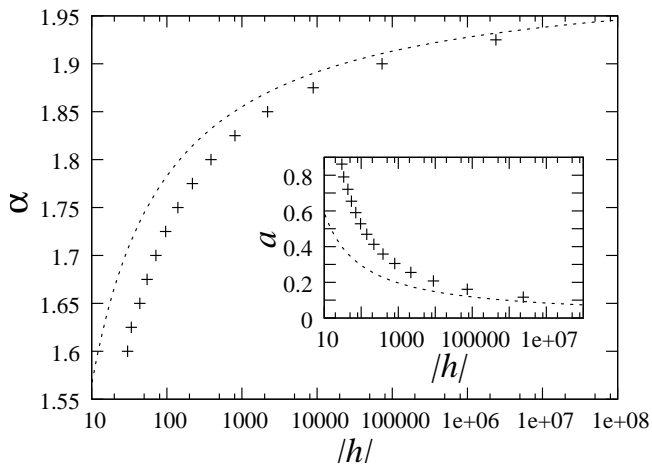}}
\end{center}
\caption{\small Optimal exponents $\alpha$ of power-laws $K_c \approx a|h|^{\alpha}$ that
approximate the scaling function $K_c(|h|)$, as shown in Fig. \ref{kcofh_fig}, as a function of $|h|$. 
The dashed curves are the corresponding asymptotic estimates of Eq. (\ref{finite_alpha}) and Eq. (\ref{finite_a}).}
\label{kc_powscale_fig} 
\end{figure}

A more detailed analysis carried out in appendix \ref{asym_kc_sec} takes into account that, for large $\bar{K}$ and $|h|$, 
according to the central limit theorem the Binomial distribution
that characterizes the damage propagation function Eq. (\ref{even_ps_eqn}) can be replaced by a Gaussian,
consequently, the expected damage is approximated very well by
\begin{equation}
\bar{d}(\bar{K},|h|) = \bar{K}\cdot \sqrt{ \frac{1}{2\pi\bar{K}}  }   \,\exp{\left[-\frac{h^2}{2\bar{K}}\right]}.
\end{equation}
Taking logarithms and inserting into Eq. (\ref{log_crit_eqn}) then yields the asymptotic scaling
\begin{equation}\label{asymp_eqn}
\lim_{|h| \to \infty} K_c(|h|) = \frac{h^2}{2\ln{|h|}}
\end{equation}
of the critical connectivity $K_c$. Figure \ref{kcofh_fig} demonstrates the convergence
of the critical line (straight-lined curve and data points) against this asymptote (dashed curve).

For finite $|h|$, we notice that there are substantial contributions
from additional terms that vanish only logarithmically, and hence an approximation based on
Eq. (\ref{asymp_eqn}) would substantially underestimate $K_c$. This can be appreciated clearly from
Fig. \ref{kerror_fig}, which demonstrates the slow (logarithmic) convergence of
the error $\varepsilon_2(|h|)$ made by application of Eq. (\ref{asymp_eqn}) for finite $|h|$.

From Fig. \ref{kcofh_fig}, it is also evident that, for finite $|h|$, Eq. (\ref{asymp_eqn}) overestimates
the slope $dK_c/d|h|$. One can show that, for finite $|h|$, $K_c(|h|)$ is
better approximated locally by power-laws of the form
\begin{equation}
K_c(|h|) \approx a(|h|)\cdot |h|^{\alpha(|h|)}
\end{equation}
with $3/2 < \alpha < 2$.
%%%%%%%%%%%%%%%%%%%%%%%%%%%%%%%%%%%%%%
We confirmed this intuition by numerically inserting
candidate solutions with fixed $\alpha$ into Eq. (\ref{log_approx_eqn}), and solving for the values of $|h|$ and $a$ where the deviation from the
true curve $K_c(|h|)$ becomes minimal; inverting this relation, we obtain the optimal power law exponents $\alpha(|h|)$
as a function of $|h|$ (Fig. \ref{kc_powscale_fig}, for details, see appendix E). Again, we can apply the Gaussian approximation
for the damage propagation function to derive upper (lower) bounds for the finite size scaling of $\alpha(|h|)$ and $a(|h|)$,
which yield (cf. appendix E)
\begin{equation}\label{finite_alpha}
\alpha(|h|) \approx 2 - \frac{1}{\ln{|h|}}.
\end{equation} 
and
\begin{equation}\label{finite_a}
a(|h|) \approx \frac{e}{2\ln{|h|}}.
\end{equation} 
Figure \ref{kc_powscale_fig} shows that the true optimal values are systematically below ($\alpha$) or above ($a$) these curves,
demonstrating the non-trivial scaling behavior of the critical line for finite $|h|$, which is significantly different from
the simple asymptotic behavior in the thermodynamic limit (Eq. (\ref{asymp_eqn})).
%%%%%%%%%%%%%%%%%%%%%%%%%%%%%%%

\begin{figure}[htb]
\begin{center}
\resizebox{85mm}{!}{\includegraphics{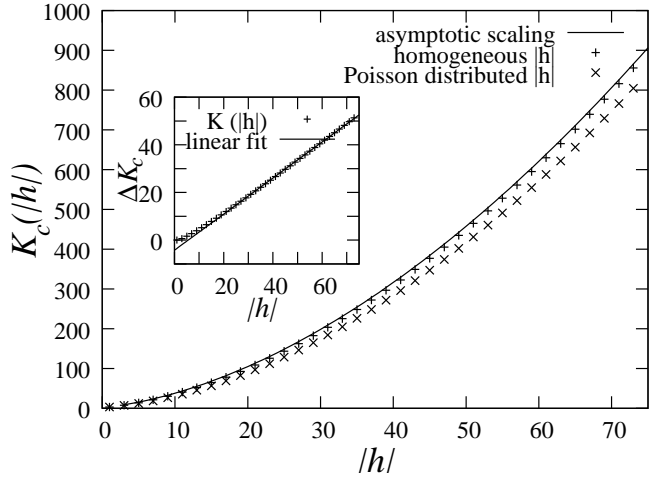}}
\end{center}
\caption{\small $K_c(|h|)$ for homogeneous thresholds (+) and Poisson distributed thresholds with the same {\em average} $\bar{|h|}$
(X), annealed approximation. The solid line is the asymptotic scaling obtained from Eq. (\ref{log_crit_eqn}). For inhomogeneous $|h|$, the critical
line is systematically below $K_c$ of networks with homogeneous $|h|$.
{\em Inset:} The difference $|\Delta K_c(|h|)|$ between both curves grows only linearly in $|h|$, confirming that the asymptotic scaling in the limit $|h| \to \infty$, is the same in both cases.}
\label{kchihfine_fig} 
\end{figure}

Let us now investigate the scaling behavior of $K_c$ for networks with inhomogeneous thresholds. 
%%%%%%%%%%%%%%%%TODOOOOOOOOOOOOOOOOOOOOOOOOOOOOOOO!
Figure \ref{kchihfine_fig}  shows that, for finite $|h|$, the critical line $K_c(|h|)$ for RTN with inhomogeneous thresholds
is always {\em below} the corresponding values for homogeneous $|h|$; the absolute {\em difference} $\Delta K_c(|h| := |K_c^{h}(|h|) - K_c^{i}(\bar{|h|} = |h|)|$
between both curves, however, increases only linearly in with $|h|$ (inset of Fig. \ref{kchihfine_fig} ), where
$K_c^{h}(|h|)$ is the critical connectivity for homogeneous $|h|$, and $K_c^{i}(\bar{|h|})$ the corresponding
value for inhomogeneously distributed  $|h|$ with mean $\bar{|h|} = |h|$.

Intuitively, this is straight-forward to understand:  
since we assumed that $k$ and $|h|$ are {\em statistically independent},
$\Delta K_c(|h|)$ is
determined solely by the {\em variance} $\sigma_h^2$ of the threshold distribution around the mean threshold $\bar{|h|} = |h|$ - the smaller this variance is,
the more peaked this distribution is around $\bar{|h|} = |h|$, and the less it hence differs from the homogeneous distribution.
Since we assumed that (in the inhomogeneous case) thresholds are Poisson distributed around $\bar{|h|}$, we directly conclude 
\begin{equation}
 \Delta K_c(|h|) \sim  \sigma_h^2 = \bar{|h|}.
\end{equation}
%%%%%%%%%%%%%%%%%%%%%%%%%%%%%%%%%%%
For arbitrary threshold distributions that are statistically independent from the networks' degree distribution
with variance $\sigma_k^2$, we make the ansatz
\begin{equation}
\sigma_{tot}^2 = \sigma_k^2 + \sigma_h^2
\end{equation}
for the total variance $\sigma_{tot}^2$. Using the same Gaussian approximation as above for the homogeneous case, one can show that
\begin{equation}
K_c(|\bar{h}|) \approx \frac{\bar{h}^2}{2\ln{|\bar{h}|}} - \sigma_h^2
\end{equation}
for networks with inhomogeneous thresholds distributed around an average absolute threshold $|\bar{h}|$ (for details, cf. appendix C).
This implies that, in the limit $|\bar{h}| \to \infty$, all networks with Poissonian distributed connectivity and threshold distributions
with a variance which obeys the scaling relation: $\sigma_h^2 \sim |\bar{h}|^{\beta}$ with $0 \le \beta < 2$, follow the universal
asymptotic scaling relation
\begin{equation} \label{uniscale1_eqn}
K_c(|\bar{h}|) = \frac{\bar{h}^2}{2\ln{|\bar{h}|}},
\end{equation}
as it is shown in appendix C.
This means that in all these cases,
the asymptotic scaling for $\bar{|h|}\to\infty$ is dominated by by the scaling behavior of the maximum
of the damage propagation function $p_s(k, |h|)$, with an exponent $\alpha = 2$.

\begin{figure}[htb]
\begin{center}
\resizebox{85mm}{!}{\includegraphics{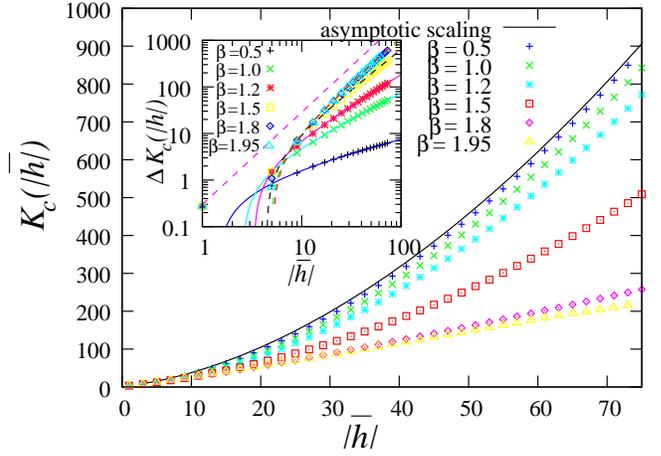}}
\end{center}
\caption{\small $K_c(\beta, \bar{|h|})$ for networks with threshold distributions following discretized Gaussian distributions with
different variances $Var(|h|) = \bar{|h|}^{\beta}$ (for details, see text). One clearly appreciates that the larger the variance of the
threshold distribution, the more the curves $K_c(\beta, \bar{|h|})$ are below the critical line of networks with homogeneous
thresholds (blue solid line); in the limiting case $\beta = 1.95 \approx \alpha$ (yellow triangles), $K_c$ 
scales almost linearly with $\bar{|h|}$. {\em Inset:} differences $|\Delta K_c(\beta, \bar{|h|})|$ to the critical line of
RTN with homogeneous thresholds scale $\sim \bar{|h|}^{\beta_{e}}$ with $\beta < \beta_{e} < \alpha$ (power law fits 
and dashed line with slope $\alpha$ shown for comparsion); this implies
asymptotic convergence to the universal scaling function Eq. (\ref{uniscale1_eqn}) in the limit $\bar{|h|} \to \infty$ for all cases shown here.    }
\label{kcbeta_fig} 
\end{figure}

\begin{table}
\begin{center}
\begin{tabular}{|  c  |  c  |}\hline
$\beta$ & $\beta_{e}$ \\ \hline \hline 
$0.5$ & $0.533 \pm 0.009$ \\ 
$1.0$ & $1.099 \pm 0.004$ \\ 
$1.2$ & $1.327 \pm 0.004$ \\ 
$1.5$ & $1.732 \pm 0.004$ \\ 
$1.8$ & $1.942 \pm 0.003$ \\ 
$1.95$ & $1.975 \pm 0.004$ \\ \hline

\end{tabular}
\end{center}
\caption{\small Scaling exponents $\beta_{e}$, as obtained from fits of $\Delta K_c \sim \bar{|h|}^{\beta_{e}}$, as
a function of $\beta$. }
\end{table}

Let us now confirm this finding for a different class of threshold distributions.
Since in a Poissonian the variance is not a free parameter, we now instead choose a discretized Gaussian distribution, i.e.
\begin{equation}
P( |h| ) = \frac{Z}{\sigma_h\sqrt{ 2\pi  }} e^{-\frac{1}{2}(|h|-\bar{|h|})^2/\sigma_h^2} 
\end{equation} 
with
\begin{equation}
Z = \left\{\sum_{|h| = 0}^{|h|_{m}} \frac{1}{\sigma_h\sqrt{ 2\pi  }} e^{-\frac{1}{2}(|h|-\bar{|h|})^2/\sigma_h^2} \right\}^{-1}
\end{equation} 
and variance
\begin{equation}
\sigma_h^2 = \bar{|h|}^\beta, \quad \beta \in [0,\alpha).
\end{equation} 
The factor $Z$ ensures that the probabilities are normalized in the interval $[0,|h|_{m}]$,
where $|h|_m$ denotes the cutoff of the threshold distribution. 
Figure \ref{kcbeta_fig}  compares the scaling functions $K_c(\bar{|h|})$ for different values of $\beta$ to the asymptotic case
of homogeneous networks. Obviously, for finite $\bar{|h|}$, increased variance of the threshold distribution
substantially lowers the critical connectivity; in the limiting case $\beta \approx \alpha$,
$K_c$ grows only linearly with $\bar{|h|}$. For $\beta < \alpha$, we find that the deviation from
the scaling behavior of RTN with homogeneous thresholds scales as
\begin{equation}
\Delta K_c \propto \bar{|h|}^{\beta_{e}}.
\end{equation} 
Table 1 compares $\beta$ and $\beta_{e}$ (as obtained from fits of $\Delta K_c$; in all cases, we have 
$\beta_{e} > \beta$, which is a discretization effect, but still $\beta_{e} < \alpha$.  Hence,
it follows that
 \begin{eqnarray}
\lim_{|h| \to \infty} \frac{K_c(\beta, \bar{|h|} = |h|)}{K_c^{h}(|h|)} &=& \lim_{|h| \to \infty} \frac{K_c^{h}(|h|) - \Delta K_c(\beta, |h|)}{K_c^{h}(|h|)} \nonumber \\
&=& 1 - const. \cdot  \lim_{|h| \to \infty}|h|^{\beta_{e} - \alpha} \\
 &=& 1 \nonumber
\end{eqnarray} 
for $\beta_{e} < \alpha$, i.e. in this case all scaling functions $K_c(\beta, \bar{|h|})$ for $|h| \to \infty$ indeed asymptotically converge to the same 
universal scaling function, as given by Eq. (\ref{uniscale1_eqn}).

\begin{figure}[htb]
\begin{center}
\resizebox{85mm}{!}{\includegraphics{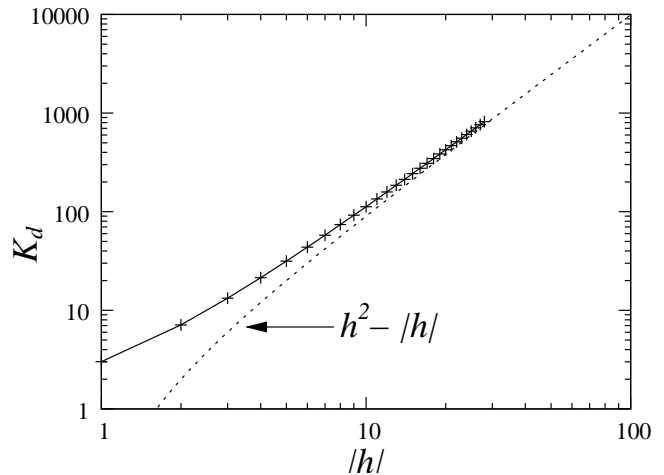}}
\end{center}
\caption{\small Scaling behavior of $K_d(|h|)$ as a function of $|h|$, double logarithmic plot. The dashed line
highlights the asymptotic scaling (Eq. (\ref{kd_asymp_eq})).}
\label{kdscalefig} 
\end{figure}

Let us now have a closer look at the scaling behavior of the intersection points $K_d(|h|)$, as introduced
in the last paragraph of subsection B. Let $\bar{d}^{h}(\bar{K}, |h|)$ be the expected damage in
networks with homogeneous threshold, and $\bar{d}^{i}(\bar{K}, \bar{|h|})$ the expected damage in
networks with inhomogeneous thresholds; then
\begin{eqnarray}
\bar{d}^{h}(K_d(|h|), |h|) -  \bar{d}^{i}(K_d(|h|), \bar{|h|}) &=& 0  \label{kd_eqn1}\\
\bar{|h|} &=& |h| \label{kd_eqn2}
\end{eqnarray} 
are the defining equations for $K_d(|h|)$. Notice that for $\bar{K} < K_d$, the randomness introduced
by inhomogeneous thresholds actually {\em increases} the probability for damage spreading, whereas
for $\bar{K} > K_d$, it is {\em decreased}. Equation (\ref{kd_eqn1}), under condition
Eq. (\ref{kd_eqn2}), can be solved numerically for not to large $|h|$. 
%%%%%%%%%%%%%%%%%%%%%%%%%%%%%%%%%%%%%%%%%%%%%
Further, one can derive the asymptotic scaling in the thermodynamic limit by application
of the Gaussian approximation for the damage propagation function (for details, cf. appendix D), showing that
\begin{equation}\label{kd_asymp_eq}
\lim_{|h| \to \infty} K_d(|h|) = h^2 - |h|.
\end{equation} 
Fig. \ref{kdscalefig}
demonstrates that $K_d(|h|)$ approaches this asymptotic scaling already for
considerably small $|h|$, indicating that $K_d(|h|)$ is characterized by the same universal scaling exponent $\alpha = 2$
as $K_c(|h|)$. Notice, however, that the asymptotic scaling law for $K_d$ obeys a purely algebraic relation,
whereas $K_c$ has a dependence $\sim h^2/\ln{|h|}$ (Eq. \ref{asymp_eqn}).

Let us briefly compare the scaling behavior of RTN with non-zero thresholds, as discussed above, to Random Boolean Networks (RBN).
Obviously, increasing $|h|$ biases the output states of network nodes (for the systems discussed in this paper, it increases
the probability to have an output state $\sigma_i = -1$).
Biased RBN obey the scaling relationship \cite{Derrida1990}
\begin{equation}\label{biased_rbn_eq}
K_c = \frac{1}{2p(1-p)}.
\end{equation} 
To compare this relationship to the asymptotic scaling for RTN in the limit of large $|h|$, we have to consider the limit $p \to 1$.
One can show that, in this limit, the scaling function Eq. (\ref{biased_rbn_eq}) logarithmically approaches the asymptotic scaling
\begin{equation}
K_c \approx -\frac{p^2}{2\ln{p}}.
\end{equation} 
This shows that $|h|$ plays the same role as the bias parameter $p$ in RBN, and that both classes obey the same scaling in the limit
$p \to 1$ and $|h| \to \infty$, respectively. However, there are also substantial differences between both classes of systems,
that come into play when $|h|$ is small (when $p$ is close to $1/2$). In particular, while RBN in this limit still obey the simple
scaling relationship Eq. (\ref{biased_rbn_eq})), the critical connectivity $K_c$ of RTN 
is derived from the complex dependence of Eq. (\ref{log_approx_eqn}). This difference is due to the fact that, in RTN,
local damage propagation strongly depends on the in-degree of nodes (cf. Eq. \ref{odd_ps_eqn} and \ref{even_ps_eqn}), while it is
independent from the in-degree in RBN for $k > 0$. In the limit of sparsely connected networks (i.e. small $|h|$ and $K_c$), this leads to much stronger finite size effects
in RTN than in RBN. Furthermore, in this limit also the absolute values of $K_c$ in RTN are considerably below those of RBN \cite{RohlfBornhol02,Nakamura2004}.

Finally, let us remark on the existence of the characteristic connectivity $K_d$. As shown above, $K_d$ is defined for
RTN with arbitrary $|h|$, in particular, it exists in the limit $|h| \to \infty$, with a well-defined asymptotic scaling.
For biased RBN, the corresponding limit is given by $p \to 1$ (or, equivalently, $p \to 0$). 
Obviously, we can in principle assign variable (inhomogeneous) biases $p_i$ to different RBN nodes 
such that the {\em average}
bias is equal to $p$. However, because $p$ is a probability and hence $0 \le p \le 1$, the variance $\sigma_p^2$ has to
vanish in the limit $p \to 1$ ($p \to 0$) to yield a proper average bias. 
Since $K_d$ is defined by comparing networks with diverging variance of
the order parameter $|h|$ (or $p$, respectively) with the corresponding networks with vanishing variance and the same
average $|h|$ (or $p$, respectively), this implies that $K_d$ {\em is not defined for RBN} in the limit of large bias $p \to 1$, which
corresponds to $|h| \to \infty$ in RTN. Hence, $K_d$ constitutes a truly novel, not previously known concept, yielding a new characteristic connectivity which is well-defined only for RTN.

It is interesting to notice that the dependence of $K_c$, as well as of $K_d$ on $|h|$ is clearly super-linear even for considerably
small $|h|$; this has profound consequences for algorithms that evolve RTN towards (self-organized) criticality by local adaptations
of both thresholds and the number of inputs a node receives from other nodes \cite{Rohlf_hevo2007}. In particular,
it can be shown that co-evolution of network dynamics and thresholds/in-degrees leads to strong correlations between
$|h|$ and $k$.
%, and broad in-degree distributions. 
%%%%%%%TODO
To approach this type of problem analytically, we will now
extend our analysis in this direction. first,
In the next section, we will show that even weak correlations between $k$ and $|h|$ can lead to a transition from sub-critical to
super-critical dynamics (and vice versa), while keeping the average connectivity $\bar{K}$ {\em and} the average absolute
threshold $\bar{|h|}$ constant.
%%%%%%%%%%%%%%%%%%%%%%%%%%%%%%%%%%%%

\begin{figure}[htb]
\begin{center}
\resizebox{85mm}{!}{\includegraphics{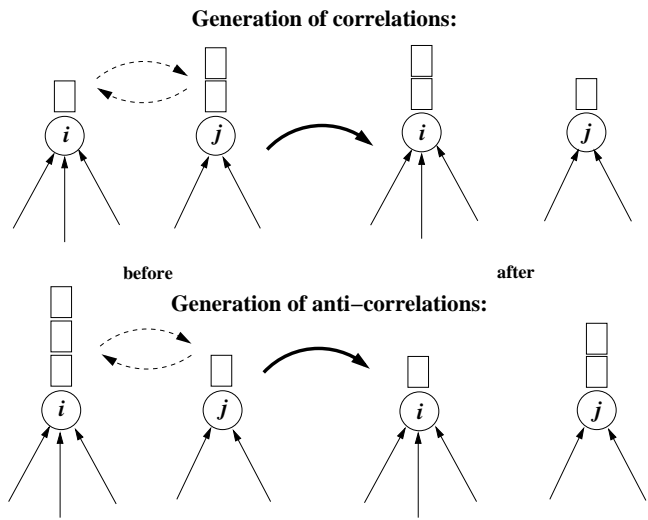}} %%fig12
\end{center}
\caption{\small Schematic illustration of the algorithm applied to generate local (anti-)correlations between in-degree
$k_{in}$ and (absolute) threshold $|h|$. Arrows symbolize inputs from other nodes, boxes symbolize node thresholds (one box corresponds
to $|h| =1$, two boxes to $|h| = 2$, and so on). For details of the algorithm, please refer to the text.}
\label{corrscheme} 
\end{figure}

\begin{figure}[htb]
\begin{center}
\resizebox{85mm}{!}{\includegraphics{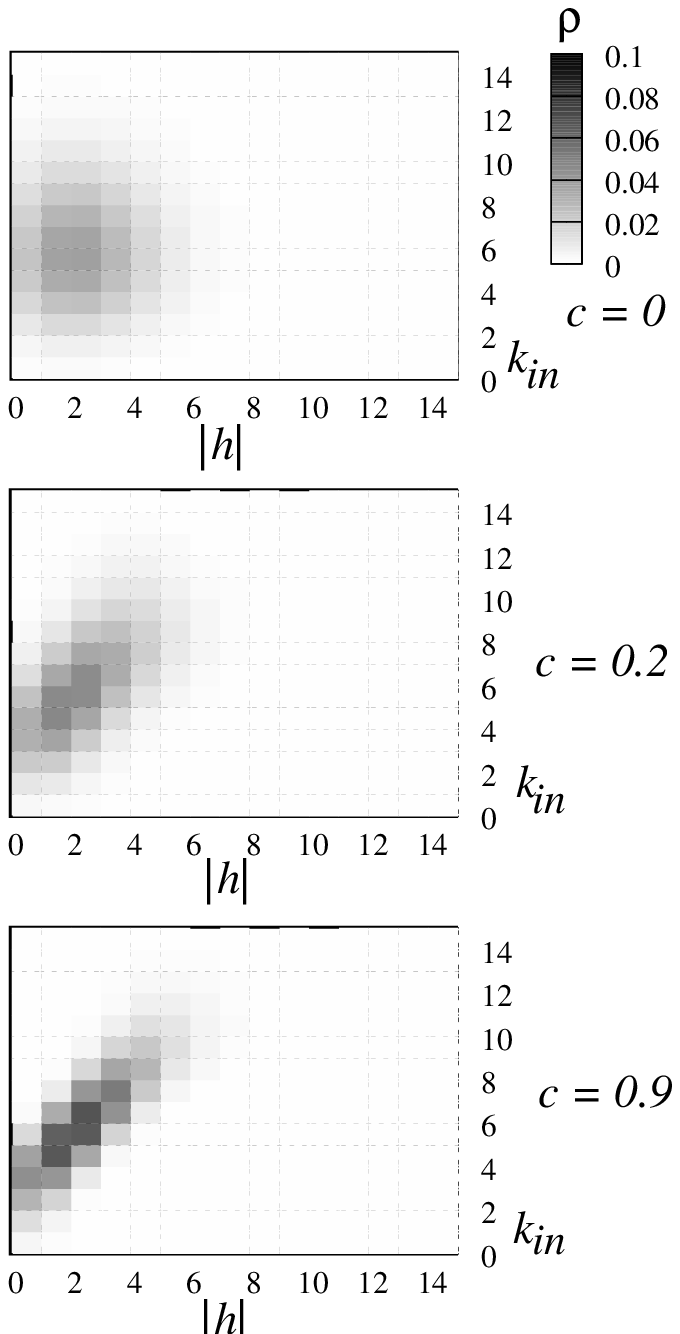}} %%fig13
\end{center}
\caption{\small Combined density $\rho(k_{in},|h|)$ for three different positive values of the correlation parameter
$c$, from top to bottom: $c = 0$, $c=0.2$ and $c = 0.9$. Dark gray indicates a high probability density. The diagonal
structure of $\rho(k_{in},|h|)$ for $c=0.9$ (lower panel) indicates emergence of strong positive correlations between
$k_{in}$ and $|h|$.}
\label{corrmatrix} 
\end{figure}

\begin{figure}[htb]
\begin{center}
\resizebox{85mm}{!}{\includegraphics{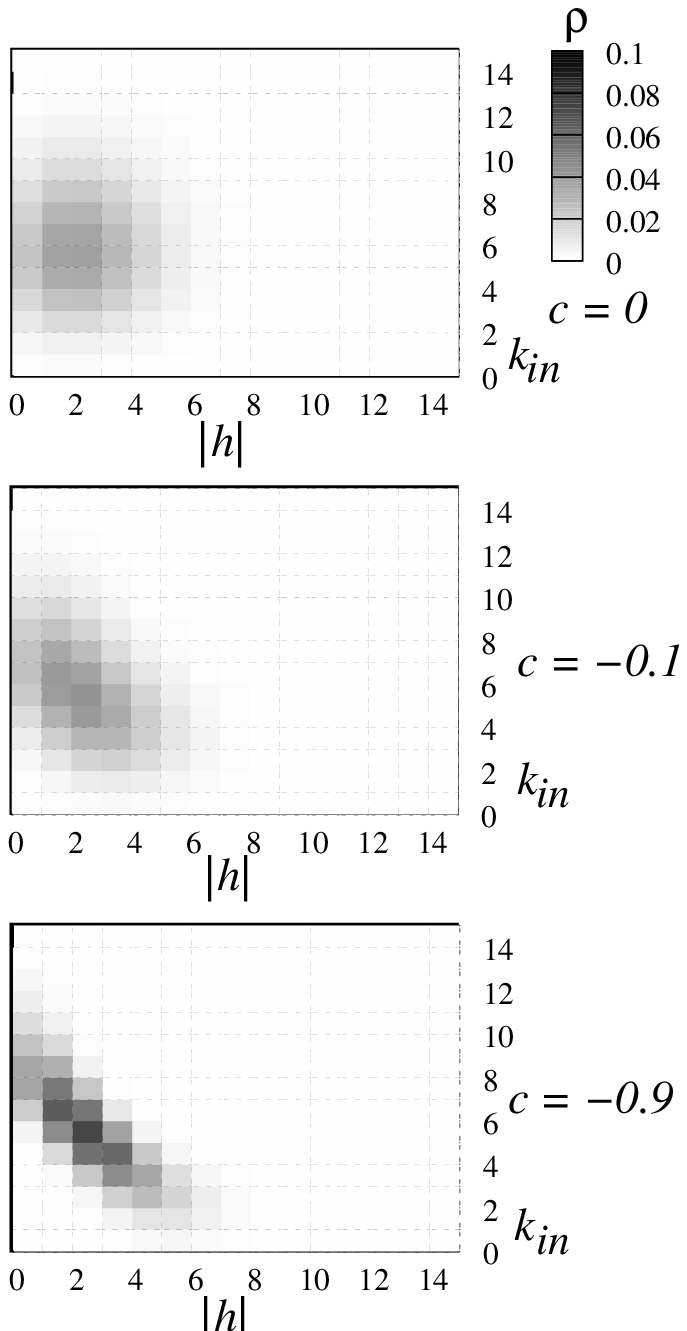}} %%fig14
\end{center}
\caption{\small Combined density $\rho(k_{in},|h|)$ for three different negative values of the correlation parameter
$c$, from top to bottom: $c = 0$, $c=-0.1$ and $c = -0.9$. Dark gray indicates a high probability density. The inverted diagonal
structure of $\rho(k_{in},|h|)$ for $c=-0.9$ (lower panel) indicates emergence of strong anti-correlations between
$k_{in}$ and $|h|$. }
\label{acorrmatrix} 
\end{figure}

\begin{figure}[htb]
\begin{center}
\resizebox{85mm}{!}{\includegraphics{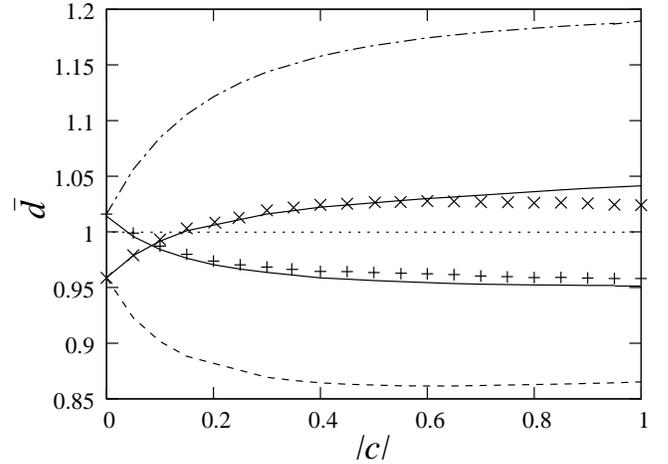}}  %%fig15
\end{center}
\caption{\small Average damage $\bar{d}(c)$ as a function of $|c|$, for correlated $k_{in}$ and $|h|$ (+) and anti-correlated
$k_{in}$ and $|h|$ (X), with $\bar{K}=6.15$ for $c \ge 0$ networks, $\bar{K}=5.8$ for $c \le 0$ networks and
$\bar{|h|} = 2.5$ in both cases. Numerical data where obtained from ensemble averages over $Z=5\cdot 10^5$
randomly generated RTN with $N=1024$ nodes for each data point. Solid curves are the corresponding results
of the {\em corrected} annealed approximation, while the dashed-dotted curve shows the uncorrected result for correlated
$_{in}k$ and $|h|$, and the dashed curve the uncorrected result for anti-correlated
$k_{in}$ and $|h|$, respectively.}
\label{hadicorrfig} 
\end{figure}

\begin{figure}[htb]
\begin{center}
\resizebox{85mm}{!}{\includegraphics{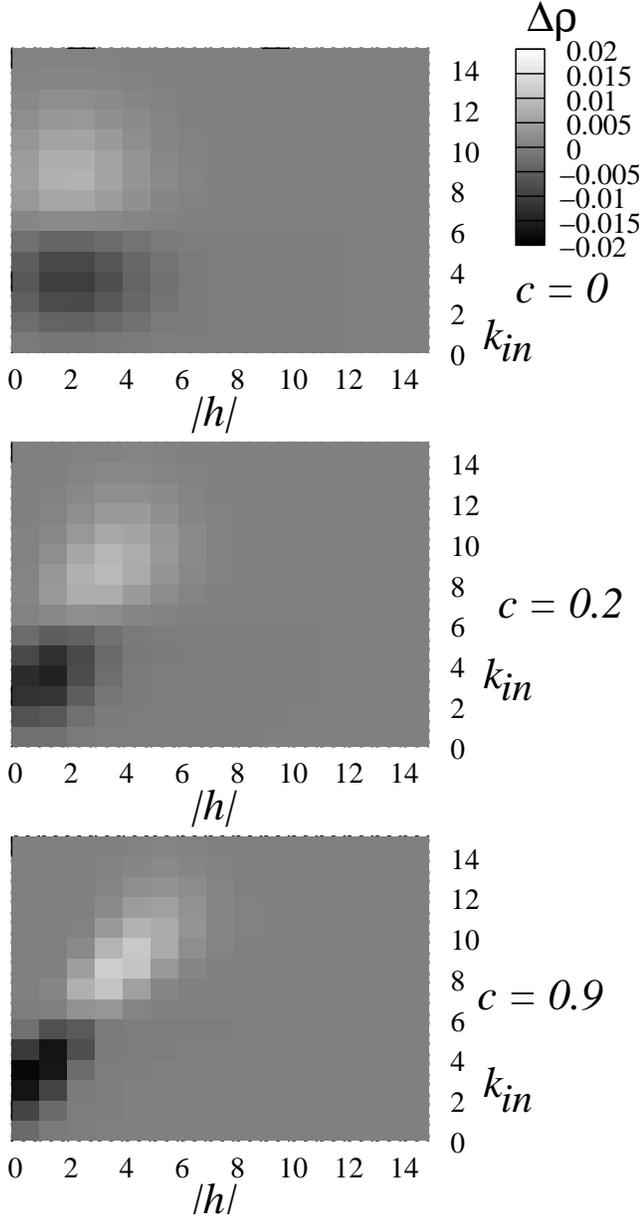}} %%fig16
\end{center}
\caption{\small Difference $\Delta\rho(k,|h|)$ between real and effective combined densities (see text),
for three different values of positive $c$. Dark gray indicates a negative deviation, light gray positive
deviation, a medium intensity refers to zero deviation.}
\label{devcorrfig} 
\end{figure}

\begin{figure}[htb]
\begin{center}
\resizebox{85mm}{!}{\includegraphics{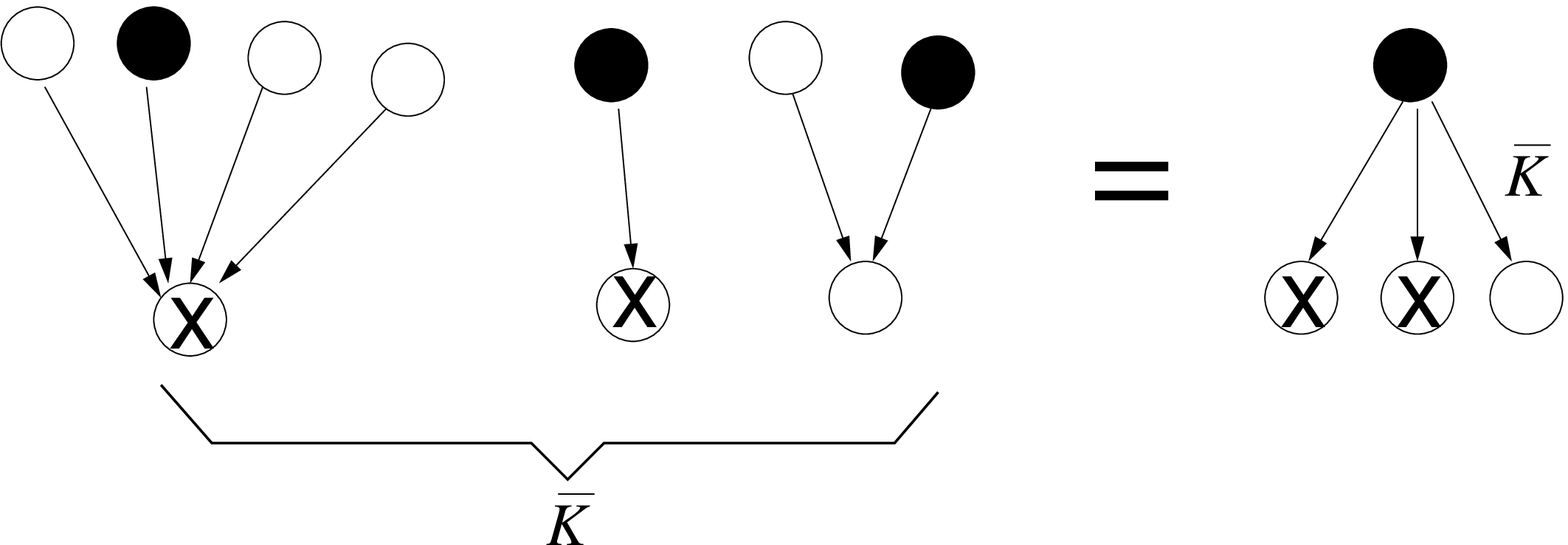}} %%fig17
\end{center}
\caption{\small Schematic illustration of the {\em naive} mean-field assumption implicit to the annealed approximation:
Choosing a random ensemble of $\bar{K}$ nodes and inverting one input of each of them 
(left panel, black circles refer to inverted states) on average yields the same damage as perturbing a randomly chosen site with $\bar{K}$ outputs and investigating the resulting
damage at the output nodes (right panel); in both panels, X means "damaged states". This assumption is violated for 
correlated networks, even if the generating algorithm {\em explicitly} correlates only in-degree and thresholds.}
\label{annealednaivefig} 
\end{figure}

\subsection{Effect of correlations between $k$ and $h$}
So far, we assumed that node degree and node thresholds are totally uncorrelated; while this matches well the "maximum disorder"
assumption used in random ensemble based approaches as, e.g.,  the annealed approximation, this might be a quite unrealistic
constraint for many real world networks. Indeed, one can show that even in a simple evolutionary algorithm that
couples both the adaptation of node thresholds $h_i$ and in-degree $k_i$ to a local dynamical order parameter, strong correlations between both quantities
emerge spontaneously \cite{Rohlf_hevo2007}. Hence, it is an interesting question to ask whether correlations (or anti-correlations) between
$h$ and $k$ may induce a transition from sub-critical to super-critical networks (or vice versa), while we keep $\bar{K}$ and
$\bar{|h|}$, and network topologies constant.

Let us first formulate an algorithm that generates correlations (anti-correlations) between $k$ and $|h|$.
For this purpose, a parameter $c \in [0,1]$ is introduced which parameterizes the probability that $k$ and $|h|$
are locally correlated (anti-correlated). The topology-generating algorithm then reads as follows (compare also Fig. \ref{corrscheme}):
\\\\
1) Generate a random, directed network with Poisson distributed $k$ and Poisson distributed $|h|$ with average connectivity $\bar{K}$ and 
$\bar{|h|}$ for all sites.\\
2) Select a pair of sites $i \le N$ and $j \le N$ at random. $c > 0$: exchange the sites' thresholds if $k_{in}(i) \ge k_{in}(j)$ and $|h|(i) \le |h|(j)$, or vice versa. $c < 0$: exchange the sites' thresholds if $k_{in}(i) \le k_{in}(j)$ and $|h|(i) \le |h|(j)$, or vice versa. \\ 
3) Go back to 2) and repeat the algorithm for $c\times P_{max}$ steps, where $P_{max}$ is a pre-defined maximum number
of correlated pairs.\\

Obviously, increasing the parameter $c \in [0,1]$ increases correlations (anti-correlations) between $k_{in}$ and $h$. If we repeat this algorithm
$Z$ times for fixed $c$, we can generate a random ensemble of $Z$ correlated/anti-correlated networks, and investigate damage spreading
on these networks. The ensemble-averaged probability $\rho(k_{in},|h|)$ to have a site with $k_{in}$ inputs and threshold $|h_i| = |h|$ then is defined as
\begin{equation}
\rho(k_{in},|h|) = \frac{\sum_{j=1}^Z n_j(k_{in},|h|)}{Z\cdot N},
\end{equation} 
where $n_j(k_{in},|h|)$ is the number of sites with  $k_{in}$ inputs and threshold $|h_i| = |h|$ in the $j$th random network.
 Figure \ref{corrmatrix} demonstrates the correlating effect of the algorithm on the {\em average} probabilities $\rho(k_{in},|h|)$
for ensembles of $10^5$ randomly generated networks, for the case $c \ge 0$, with $P_{max} = 10^4$. For $c = 0$, clearly no correlations are present,
and the combined density simply represents the independent superposition of the two underlying Poisson distributions. With increasing
$c$, correlations gradually emerge, and for $c = 0.9$ the resulting distribution clearly exhibits a diagonal structure. Figure \ref{acorrmatrix} demonstrates the corresponding effect for $c \le 0$, i.e. anti-correlated topologies.

Let us now investigate how these correlations affect damage propagation. To apply the annealed approximation, we now have to
calculate the average probability for damage propagation (in a finite network of size $N$) according to
\begin{equation}
\langle p_s\rangle(\bar{K},\bar{|h|},c) 
= \sum_{|h|=0}^{|h|_{m}}\sum_{k=|h|}^N \rho(k_{in},|h|)\,p_s(k+1,|h|),\label{pscorreq}
\end{equation}
with the normalization conditions
\begin{equation}
\sum_{|h|=0}^{|h|_{m}}\sum_{k=|h|}^N \rho(k_{in},|h|) = 1
\end{equation}
and 
\begin{equation}
\sum_{|h|=0}^{|h|_{m}}\sum_{k=|h|}^N |h|\,\rho(k_{in},|h|) = \bar{|h|},
\end{equation}
where $|h|_{m}$ is the maximal absolute threshold observed (cutoff);
correlations enter via the probabilities $\rho(k_{in},|h|)$ to observe a node with in-degree
$k_{in}$ and absolute threshold $|h|$. 

Figure \ref{hadicorrfig} compares the numerically observed damage $\bar{d}$ (for ensembles of randomly generated
networks) one time step after a one-bit perturbation
to the expected damage $\langle p_s\rangle(\bar{K},\bar{|h|},c)\cdot \bar{K}$, as predicted by the annealed
approximation (dashed curves). In both cases, for correlations and anti-correlations, the so-obtained
theoretical curves are obviously wrong both with regard to quantitative matching of the numerical data,
as with regard to the predicted trend: while in the numerical experiment a decrease of $\bar{d}$ with
increasing $c > 0$ is observed, i.e. a transition from super-critical (chaotic)
to sub-critical (ordered) dynamics, the annealed curve predicts a strong increase. Corresponding observations
(with opposite signs) are made for the case of anti-correlations. We conclude that there must be an additional
effect present which is not captured in our {\em naive mean-field model}. To identify the origin of this deviation,
one has to compare the distribution $\tilde{\rho}(k_{in},|h|)$ which is observed for the {\em outputs of pertubed sites}
with the original distribution $\rho(k_{in},|h|)$, which is averaged over the whole topology. 
Figure \ref{devcorrfig} shows the deviation $\Delta \rho(k_{in},|h|) := \tilde{\rho}(k_{in},|h|) - \rho(k_{in},|h|)$ between both distributions. While for small $k_{in}$ and $|h|$ negative deviations are found, for larger  $k_{in}$ and $|h|$ deviations have a
positive sign. One can easily understand the source of this {\em effective bias}, when one thinks of the correlation-generating
mechanism (Fig. \ref{corrscheme}): if we pick, by chance, within a correlated network a site $i$ with small $k_{in}$,
it will probably also have a small $|h|$, and its outputs will probably have {\em larger} $k_{in}$ and $|h|$ than site $i$.
Since both $k_{in}$ and $|h|$ are bounded from below, this leads to a systematic bias to observe larger $k_{in}$ and $|h|$,
at the expense of smaller, for the outputs of perturbed nodes. The opposite effect obviously holds for sites with
large $k_{in}$ and $|h|$. While for $c = 0$ this effect is very small, it becomes dominant for $c \to 1$ due to the strong asymmetry
of the diagonal distribution. If we correct the annealed approximation to include this bias, by replacing $\rho(k_{in},|h|)$
with $\tilde{\rho}(k_{in},|h|)$ in Eq. (\ref{pscorreq}), we find that the resulting {\em corrected} annealed curves much better
match the numerically observed damage (still, there are slight discrepancies for large $c$, which are due to finite ensemble sizes).

The result of this study demonstrates that the annealed approximation has to be used with extreme care, when topological
correlations are present. Although the applied algorithm {\em explicitly} correlates only in-degree and thresholds, consideration of
these correlations only by using a combined density  $\rho(k_{in},|h|)$ for averages over topology and dynamics
leads to wrong predictions. In addition, one has to consider systematic bias effects between perturbed sites and their outputs,
which arise as a side effect of the correlating algorithm. This shows that the naive mean-field assumption inherent to the annealed
approximation, as demonstrated in Fig. \ref{annealednaivefig}, is violated even in the simplest case of correlations between
in-degree and thresholds. Instead, complete information about topology, including the output side, has to enter
statistics, and the order of averages becomes important.

\section{Discussion}
An increasing number of studies is concerned with the propagation dynamics of perturbations and/or information
in complex dynamical networks. Discrete dynamical networks, in particular Random Boolean Networks (RBN) and
Random Threshold Networks (RTN), constitute an ideal testbed for this type of question, since they are
easily accessible for both computational methods and the tool boxes of statistics and combinatorics. 
Often, it is found that damage/information propagation strongly depends on the type of inhomogeneities present in
network wiring. Several studies focus, for example, on the effect of scale-free degree distributions
\cite{Nakamura2004,Aldana2004}. Typically, these studies employ mean-field methods
and hence represent, in a sense, strongly idealized models, since they derive results that strictly hold
in the thermodynamic limit only.

Consequently, a second line of research concentrates on modification
of damage propagation due to finite-size effects, which play a decisive role in many real-world
networks. Recently, it was shown that weakly perturbed, finite size RBN and RTN show pronounced
deviations from the annealed approximation \cite{RohlfGulbahceTeuscher2007}. Fronczak and Fronczak showed that these deviations
can be explained by inhomogeneities and emergent correlations found at the percolation transition \cite{Fronczak08},
however, their study is currently limited to undirected networks. In this context, the system
discussed in our paper constitutes a complementary approach: it allows to introduce {\em dynamical
inhomogeneity} of network units, without otherwise altering network topology. While this type
of dynamical diversity certainly plays an important role in many real-world networks, it is neglected
by most researchers. Let us now briefly summarize the main results of our study.

We studied damage propagation in Random Threshold Networks (RTN) with homogeneous and inhomogeneous
negative thresholds, both analytically (using an annealed approximation) and in numerical simulations.
We derived the probability $p_s(k,|h|)$ of damage propagation for arbitrary in-degree $k$ and (absolute) threshold $|h|$
(Eqs. (5)-(8)), and, from this, the corresponding {\em annealed} probabilities $\langle p_s\rangle$ (Eq. (10) and Eq. 12)) and the expected damage
$\bar{d}$ (Eq. (11)), for both the cases of homogeneous and inhomogeneously distributed thresholds.
On these grounds, we investigated the scaling behavior of the critical connectivity $K_c$ as a function of $|h|$.
Using a mean field approximation, a simplified scaling equation for the logarithm of the average damage was derived (Eq. (14)),
and applied to derive the critical line $K_c(|h|)$ (Fig. 6). It was shown that this function exhibits a super-linear
increase with $|h|$, which asymptotically approaches a unique scaling law $K_c(|h|) \sim h^2/(2 \ln{|h|})$ for large $|h|$
(Eq. (18) and Fig. 7). However, convergence against this asymptotic scaling is very slow (logarithmic in $|h|$),
which indicates that finite size effects are very dominant, and cannot be neglected for realistically sized
networks. We presented evidence that this asymptotic scaling  is universal for RTN with Poissonian distributed connectivity and
threshold distributions with a variance that grows slower than $h^2$, for both the cases of Poisson distributed
thresholds (Fig. 8) and thresholds distributed according to a discretized Gaussian (Fig. 9). 
Interestingly, inhomogeneity in thresholds, meaning that each site has an individual threshold $|h_i|$ drawn, e.g., from
a Poisson distribution with mean $\bar{|h|}$, increases damage for small average connectivity $\bar{K}$, when compared
to homogeneous networks with the same average threshold $|h| = \bar{h}$,  whereas for larger $\bar{K}$ with $\bar{K} > K_d$,
damage is reduced. This establishes a new characteristic connectivity $K_d(|h|)$ with $K_d > K_c$, that describes the ambivalent effect
of threshold inhomogeneity on RTN dynamics. We showed that $K_d(|h|)$ asymptotically converges against a unique scaling law
$K_d \sim h^2$ in the limit $|h| \to \infty$. 
The scaling of $K_d$ was compared to the corresponding case of random Boolean networks (RBN) with inhomogeneously distributed bias,
parameterized in terms of a bias parameter $1/2 \le p \le 1$. It was shown that
$K_d$ {\em is not defined for RBN} in the limit $p \to 1$, which
corresponds to $|h| \to \infty$ in RTN. Hence, $K_d$ constitutes a truly novel, not previously known concept, yielding a new characteristic connectivity which is well-defined only for RTN.

Last, we introduced local correlations between in-degree $k_{in}$ of network nodes and their (absolute) threshold $|h|$,
while keeping all
other network parameters constant. We found that even small positive correlations can induce a transition from
supercritical (chaotic) to subcritical (ordered) dynamics, while anti-correlations have the opposite effect.
It was shown that the naive mean-field assumption typical for the annealed approximation leads to false predictions
in this case. Even in the simplest case, where only in-degree and (absolute) threshold are correlated,
complete information about topology, including the output side, has to enter
statistics, and the order of averages becomes important.

To summarize, dynamics of damage (or information) propagation in RTN with inhomogeneous thresholds and Poisson distributed connectivity
shows both similarities and differences, when compared to networks with homogeneous thresholds: similarities manifest themselves
in common universal scaling functions for both $K_c$ and $K_d$, whereas differences show up in the opposite effects of threshold inhomogeneity
for small and large $\bar{K}$. Differences become even more prominent in networks that are characterized by correlations between in-degree and thresholds. In this case, the annealed approximation has to be used with extreme care.
Many dynamical systems in nature, that can be described
as complex networks, exhibit
considerable variation of activation thresholds among the elements they consist of, however, these variations are often neglected
(e.g., in Boolean network based models of gene regulation networks). Our results indicate that, while general characteristics as,
for example, the scaling behavior of critical points, may be conserved in approxmations of this type, inhomogeneous thresholds can strongly impact the details
of network dynamics, and hence should be taken into account in models that aim to give a realistic description of the dynamics of, e.g.,  
gene regulation networks. 

\section{Acknowledgments}
The author thanks A. H\"ubler for interesting discussions and careful reading of the manuscript, and acknowledges
significant contributions of an anonymous referee with regard to the discussion of scaling behavior.

\appendix
\section{Derivation of $p_s(k,|h|)$}
In this section, we provide a derivation of the local damage propagation probability $p_s(k,|h|$.

Consider a network site $i$ with $k$ inputs; $k_+$ of these have positive sign, $k_{-}$ negative sign, hence,
$k_+ + k_{-} = k$. We no derive the conditions under which a inversion of one input spin at time $t$ leads to a switch
of the output of site $i$ at time $t+1$.

{\bf 1) $k-|h|$ odd:} From Eqs.  \ref{sign_eqn} %threshold sum
and  \ref{f_i_eqn}%sign function
it is easy to see that input-spin flips produce "damage" only if one of the following conditions holds:
\begin{equation}
k_+ - k_{-} - |h| = 1 \label{odd1}
\end{equation}
or 
\begin{equation}
k_+ - k_{-} - |h| = -1. \label{odd2}
\end{equation}
In case \ref{odd1}, only the reversal of positive spins is effective, whereas in case \ref{odd2}, only the reversal
of negative spins has an effect. We have
\begin{equation}
k_+ = \frac{k+|h|+1}{2} \label{odd3}
\end{equation}
in the first case and
\begin{equation}
k_{-} = \frac{k-|h|+1}{2} \label{odd4}
\end{equation}
in the second case. There is a total number of $k\cdot 2^k$ possible spin configurations, of which $\binom{k}{(k+|h|+1)/2}$ fulfill condition
\ref{odd3} and $\binom{k}{(k-|h|+1)/2}$ fulfill condition \ref{odd4}. Hence, the damage propgation probability follows as
\begin{eqnarray}
p_s(k,|h|) &=& k^{-1}\cdot 2^{-(k+1)}\cdot\left[ (k+|h|+1)\cdot \binom{k}{\frac{k+|h|+1}{2}}\right. \nonumber\\
& &\left. + (k-|h|+1)\cdot\binom{k}{\frac{k-|h|+1}{2}}\right] \\
&=& \frac{2^{-(k-1)}(k-1)!}{[(k+|h|-1)/2]![(k-|h|-1)/2]!}\\
&=& 2^{-(k-1)}\binom{(k-1)}{\frac{k+|h|-1}{2}}.
\end{eqnarray}
{\bf 2) $k-|h|$ even:} Here, we have as necessary conditions
\begin{equation}
k_+ - k_{-} - |h| = 0 \label{even1}
\end{equation}
or 
\begin{equation}
k_+ - k_{-} - |h| = 2. \label{even2}
\end{equation}
In the first case, only the reversal of negative spins is effective, whereas in the latter case the same holds for positive spins. We have
\begin{equation}
k_{-} = \frac{k-|h|}{2} \label{even3}
\end{equation}
in the first case and
\begin{equation}
k_{+} = \frac{k+|h|+2}{2} \label{even4}
\end{equation}
in the second case. 
There is a total number of $k\cdot 2^k$ possible spin configurations, of which $\binom{k}{(k-|h|)/2}$ fulfill condition
\ref{even3} and $\binom{k}{(k+|h|+2)/2}$ fulfill condition \ref{even4}. Hence, the damage propgation probability follows as
\begin{eqnarray}
p_s(k,|h|) &=& k^{-1}\cdot 2^{-(k+1)}\cdot\left[ (k-|h|)\cdot \binom{k}{\frac{k-|h|}{2}}\right. \nonumber\\
& &\left. + (k+|h|+2)\cdot\binom{k}{\frac{k+|h|+2}{2}}\right] \\
&=& \frac{2^{-(k-1)}(k-1)!}{[(k-|h|-2)/2]![(k+|h|)/2]!}\\
&=& 2^{-(k-1)}\binom{(k-1)}{\frac{k+|h|}{2}}.
\end{eqnarray}

\section{Derivation of the scaling equation}
For RTN with Poisson distributed in- and out-degree, the critical line is given by the condition
\begin{equation}
\bar{d}(t+1) = \langle p_s\rangle(K_c(|h|),|h|)\cdot K_c(|h|) = 1. \label{critdef_eqn_appB}
\end{equation}
with
\begin{equation}
\langle p_s\rangle(\bar{K},|h|) = e^{-\bar{K}}\sum_{k=|h|}^N \frac{\bar{K}^k}{k!} \,p_s(k+1,|h|). \label{ER_spread_eqn_appB}
\end{equation}
Instead of averaging over the ensemble of all possible network topologies as in Eq. (\ref{ER_spread_eqn_appB}), we now
make an explicit {\em mean field approximation}, and consider a "typical" network node with $k \approx \bar{K}$ inputs. Consequently, we 
approximate
\begin{equation}
\langle p_s\rangle(\bar{K},\bar{|h|}) \approx p_s( \lfloor \bar{K}\rfloor, |h|),
\end{equation}
where $\lfloor .\rfloor$ denotes the floor function.
In the limit of large $\bar{K}$ and $|h|$, the difference between the damage propagation probabilities for even and odd $k$ vanishes, i.e. we can set
\begin{equation} \label{approx_ps_eqn_appB}
\langle p_s\rangle(\bar{K},\bar{|h|}) \approx 2^{-(\lfloor \bar{K}\rfloor-1)}\binom{(\lfloor \bar{K}\rfloor-1)}{\frac{\lfloor \bar{K}\rfloor+|h|}{2}},
\end{equation}
and hence
\begin{equation} \label{approx_dam_eqn_appB}
\bar{d}(\bar{K},|h|) = \bar{K}\cdot 2^{-\lfloor \bar{K}\rfloor}\binom{\lfloor \bar{K}\rfloor}{\frac{\lfloor \bar{K}\rfloor+|h|}{2}}
\end{equation}
without loss of generality.

Using the Stirling approximation 
$n! \approx n^n e^{-n} \sqrt{2\pi n}$,
dropping the floor function
(since we now consider a function of real-valued variables only) and taking logarithms, we obtain
\begin{equation}
\ln{[\bar{d}(\bar{K},|h|)]} \approx \ln{\bar{K}} - \ln{2}\cdot\bar{K} + Z_1 - Z_2 - Z_3
\end{equation}
with
\begin{equation*}
Z_1 = \ln{[\bar{K}^{\bar{K}}e^{-\bar{K}}\sqrt{2\pi\bar{K}}]},
\end{equation*}
\begin{equation*}
Z_2 = \ln{ \left[\left(\frac{\bar{K}-|h|}{2}\right)^{\frac{\bar{K}-|h|}{2}}e^{-\frac{\bar{K}-|h|}{2}}\sqrt{\pi(\bar{K}-|h|)}\right] }
%&&{}\cdot\left.
%\sqrt{\pi(k-|h|-2)}\right] }
\end{equation*}
and
\begin{equation*}
Z_3 = \ln{\left[\left(\frac{\bar{K}+|h|}{2}\right)^{\frac{\bar{K}+|h|}{2}}e^{-\frac{\bar{K}+|h|}{2}}\sqrt{\pi(\bar{K}+|h|)}\right]}
\end{equation*}

Summing out the logarithms in $Z_1$, $Z_2$ and $Z_3$, one realizes that all terms linear in $\bar{K}$ drop out, resulting in
\begin{eqnarray}
\ln{[\bar{d}(\bar{K},|h|)]} &\approx& \ln{\bar{K}} + \left(\bar{K}-\frac{1}{2}\right)\ln{\bar{K}}  \nonumber \\
&& - \frac{\bar{K}-|h|+1}{2}\ln{(\bar{K}-|h|)} \nonumber \\
&& -\frac{\bar{K}+|h|+1}{2}\ln{(\bar{K}+|h|)} + C
\end{eqnarray}
with $C = \ln{\left(\sqrt{2/\pi}\right)}$.
Using some simple algebra and approximating $|h|+1 \approx |h|$, this can be reformulated as
\begin{eqnarray}
\ln{[\bar{d}(\bar{K},|h|)]} &\approx&  \ln{\bar{K}} -\frac{1}{2}\left\{ \ln{(\bar{K}} \right.\nonumber \\
&& - \bar{K}\ln{\left[\frac{(\bar{K}+|h|)(\bar{K}-|h|)}{\bar{K}^2}\right]}  \nonumber \\
&& \left. + |h|\ln{\left[\frac{\bar{K}+|h|}{\bar{K}-|h|}  \right] }\right \} + C.
\end{eqnarray}
This leads to the final result
\begin{eqnarray}
\ln{[\bar{d}(\bar{K}, |h|)]} \approx && \left.\frac{1}{2}\,\right\{\ln{\bar{K}}%\nonumber\\
-\bar{K}\cdot\ln{\left[1-\left(\frac{|h|}{\bar{K}}\right)^2  \right]} \nonumber \\
&-&\left.|h|\ln{\left[\frac{\bar{K}+|h|}{\bar{K}-|h|}\right]}\right\} + C.
\end{eqnarray}

\section{Asymptotic scaling of $K_c$} \label{asym_kc_sec}
Let us now derive the asymptotic scaling behavior of the critical connectivity $K_c(|h|)$. We start with the
case of homogeneous thresholds, and then generalize to inhomogeneous thresholds.

First, we note that the right hand-side of Eq. (\ref{approx_dam_eqn_appB}) has the form of a Binomial distribution
\begin{equation}
P(n,k) = \binom{n}{k}p^n q^{n-k}
\end{equation}
with $p = q = 1/2$, $n = \lfloor \bar{K}\rfloor$ and $k = (\lfloor \bar{K}\rfloor+|h|)/2$, multiplied with 
a prefactor $\bar{K}$. In the limit $\bar{K} \to \infty$ and $|h| \to \infty$, we can replace the Binomial distribution with
a Gaussian and drop the floor function, i.e.
\begin{eqnarray} 
\bar{d}(\bar{K},|h|) &=& \bar{K}\cdot C_n\,\exp{\left[-\frac{\left(\frac{\bar{K}+|h|}{2}-\frac{\bar{K}}{2}\right)^2}{2\bar{K}\cdot \frac{1}{2}\cdot \frac{1}{2} }\right]}.
\end{eqnarray}
This simplifies to 
\begin{eqnarray} \label{gauss_dam_eqn_appC}
\bar{d}(\bar{K},|h|) &=& \bar{K}\cdot \sqrt{ \frac{1}{2\pi\bar{K}}  }   \,\exp{\left[-\frac{h^2}{2\bar{K}}\right]}
\end{eqnarray}
with the normalization constant $C_n = \sqrt{ 1/(2\pi\bar{K}) }$ and variance $\sigma^2 = \bar{K}$.

In the case of inhomogeneous thresholds, we can still use this approximation, however, the variance $\sigma_h^2$ of the threshold distribution
adds to the variance of the damage propagation function of the homogeneous case. This implies that we have to replace $\bar{K}$
with $\bar{K}+\sigma_h^2$, and hence
\begin{eqnarray} \label{gauss_daminhom_eqn_appC}
\bar{d}(\bar{K},|\bar{h}|) &=& \frac{\bar{K}+\sigma_h^2}{\sqrt{2\pi(\bar{K}+\sigma_h^2) }  }   \,\exp{\left[-\frac{\bar{h}^2}{2(\bar{K} + \sigma_h^2)}\right]}.
\end{eqnarray}
To obtain the criticality condition, we take logarithms and set the result to zero, leading to
\begin{equation}
\ln{[K_c+\sigma_h^2]} -\frac{1}{2}\ln{[2\pi(K_c+\sigma_h^2)]} - \frac{\bar{h}^2}{2(K_c+\sigma_h^2)} = 0.
\end{equation}
This simplifies to
\begin{equation}\label{approx_h_eq}
\bar{h}^2 = (K_c+\sigma_h^2)\ln{\left[\frac{K_c+\sigma_h^2}{2\pi}\right]}.
\end{equation}
To solve this equation with respect to $K_c$, we make the ansatz
\begin{equation}\label{asym_ansatz}
K_c+\sigma_h^2 \approx \frac{\bar{h}^2}{2\ln{|\bar{h}|}}.
\end{equation}
Inserting for $K_c+\sigma_h^2 $ into Eq. (\ref{approx_h_eq}), we obtain
\begin{eqnarray} 
\bar{h}^2 &\approx& \frac{\bar{h}^2}{2\ln{|\bar{h}|}} \ln{\left[\frac{\bar{h}^2}{4\pi\ln{|\bar{h}|}}  \right]} \\
 &=& \bar{h}^2\left\{ 1 - \frac{\ln{[4\pi\ln{|\bar{h}}]}}{2\ln{|\bar{h}|}}   \right\}.\label{finite_eqn_appC}
\end{eqnarray}
Since the second term in the bracket vanishes logarithmically for $|\bar{h}| \to \infty$, we have verified that 
Eq. (\ref{asym_ansatz}) yields the correct asymptotic scaling. Consequently, the asymptotic scaling
of the critical line for large $|\bar{h}|$ is given by
\begin{equation}
K_c(|\bar{h}|) \approx \frac{\bar{h}^2}{2\ln{|\bar{h}|}} - \sigma_h^2.
\end{equation}
However, notice that the convergence is very slow, as can be appreciated from the logarithmic finite-size term in
Eq. (\ref{finite_eqn_appC}).
In particular, we conclude that the asymptotic scaling for networks with homogeneous thresholds, i.e. $|h| = const.$ and $\sigma_h = 0$ is given by
\begin{equation}
K_c^{hom}(|h|) \approx \frac{h^2}{2\ln{|h|}}.\label{hom_eq}
\end{equation}
Let us now prove that this scaling is universal 
for $|\bar{h}| \to \infty$ for all threshold distributions possessing a variance $\sigma_h^2 \sim |\bar{h}|^{\alpha}$ 
with $0 \le \alpha < 2$.
In this case, we have
\begin{eqnarray} 
\lim_{|h| \to \infty} \frac{K_c(|\bar{h}|)}{K_c^{hom}(|h|)} &=& \lim_{|h| \to \infty} \frac{K_c^{hom}(|h|) - \sigma_h^2}{K_c^{hom}(|h|)}\\
&=&  1 - \lim_{|h| \to \infty}\frac{\sigma_h^2}{K_c^{hom}(|h|)}\\
&=& 1 - \lim_{|h| \to \infty}\frac{2\ln{|h|}\cdot |h|^{\alpha}}{h^2}\\
&=&  1 - \lim_{|h| \to \infty} \frac{2\ln{|h|}}{|h|^{2-\alpha}}. \label{uni_eq}
\end{eqnarray}
Since we assumed $0 \le \alpha < 2$, the limit in Eq. (\ref{uni_eq}) vanishes, and hence the asymptotic scaling equation
\ref{hom_eq} is indeed universal for this class of threshold distributions.

\section{Asymptotic scaling of $K_d$}
The characteristic connectivity $K_d$ is defined by the conditions: 
\begin{equation}
|h| = |\bar{h}|, 
\end{equation}
where $|h|$ is the (constant)
threshold of a homogeneous network, and $|\bar{h}|$ the average threshold of a corresponding network with inhomogeneous
thresholds, and
\begin{equation}
\bar{d}^{h}(K_d(|h|), |h|) -  \bar{d}^{i}(K_d(|h|), \bar{|h|}) = 0, 
\end{equation}
where $\bar{d}^{h}$ is the expected damage for homogeneous networks, and $\bar{d}^{i}$ is the expected damage for inhomogeneous networks.
Let us further assume that thresholds are Poissonian distributed, i.e. $\sigma_h^2 = |h|$.
If we apply the same Gaussian approximation as in section \ref{asym_kc_sec}, these conditions lead to
\begin{equation}
\frac{e^{-h^2/(2K_d)}}{\sqrt{2\pi K_d}} = \frac{e^{-h^2/(2(K_d+|h|))}}{\sqrt{2\pi (K_d+|h|)}}.
\end{equation}
Taking logarithms and reordering, this reduces to
\begin{equation}
\ln{\left[\frac{K_d+|h|}{K_d}\right]} - \frac{h^2}{K_d} + \frac{h^2}{K_d+|h|} = 0
\end{equation}
Linearization of the first term leads to the approximation
\begin{equation}
\frac{|h|}{K_d} - \frac{h^2}{K_d} + \frac{h^2}{K_d+|h|} \approx 0.
\end{equation}
Solving this equation for $K_d$ finally yields the asymptotic scaling
\begin{equation}
K_d(|h|) \approx h^2 - |h|,
\end{equation}
i.e. $K_d$ scales quadratically with $|h|$.

\section{Power-law approximation of $K_c(|h|)$ for finite $|h|$}

\begin{figure}[htb]
\begin{center}
\resizebox{85mm}{!}{\includegraphics{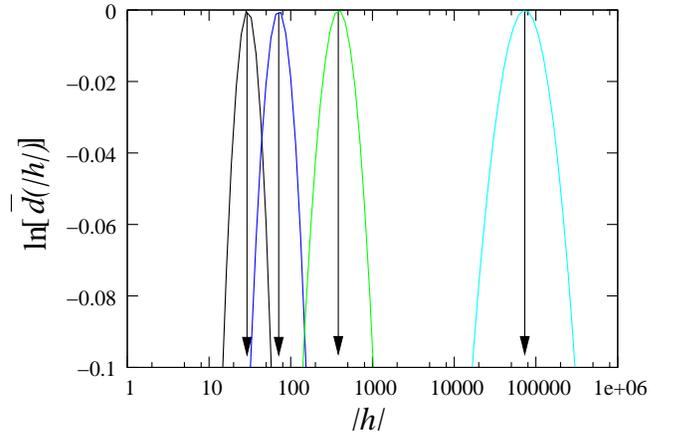}}
\end{center}
\caption{\small Solutions of Eq. (\ref{opt_alpha_eqn}) for (from the left to the right) $\alpha = 1.6$,  $\alpha = 1.7$,  $\alpha = 1.8$ and  $\alpha = 1.9$.
Projections of the maximum on the $|h|$-axis (as indicated by arrows) yield the corresponding values of $|h|_c$ at which the approximations are optimal.}
\label{optpowfit_fig} 
\end{figure}

In this section, we first describe how to identify numerically candidate solutions (power-laws)
\begin{equation}
K_c(|h|) \approx a(|h|)\cdot |h|^{\alpha(|h|)}
\end{equation} 
that optimally approximate Eq. (\ref{log_approx_eqn}) for finite (critical) $|h|_c$.

We start with a fixed $\alpha \in [1.6,2)$ and define
\begin{eqnarray}
F(y) && := \left.\frac{1}{2}\,\right\{\ln{y}%\nonumber\\
-y\cdot\ln{\left[1-\left(\frac{|h|}{y}\right)^2  \right]} \nonumber \\
&-&\left.(|h|+1)\ln{\left[\frac{y+|h|}{y-|h|}\right]}\right\} + C
\end{eqnarray}
with $y = a\cdot |h|^{\alpha}$.
One can show that, for any finite $a$ and $\alpha$, $F(y)$ has a maximum at a finite value $|h|_{max}$.
We know that $K_c$ is a monotonically increasing function of $|h|$, and intend to optimize the power-law approximation
exactly at $K_c$. Hence, we  have to vary $a$ such that
\begin{equation}
\max_{a} F(y) |_{\alpha} = 0. \label{opt_alpha_eqn}
\end{equation} 
Projection of the maximum on the $|h|$-axis then yields the corresponding threshold values $|h|_{c}(\alpha)$ at which the approximation
for the given $\alpha$ is optimal (Fig. \ref{optpowfit_fig} ). Inversion of this relation allows us to plot 
the corresponding values of the function $\alpha(|h|)$ (Fig. \ref{kc_powscale_fig}).

Last, let us estimate the asymptotic scaling of $\alpha(|h|)$. If we apply the asymptotic scaling relation for $K_c$ derived
in section \ref{asym_kc_sec}, we can approximate
\begin{equation}
\frac{h^2}{2\ln{|h|}} = a(|h|)\cdot |h|^{\alpha(|h|)}.
\end{equation} 
Taking logarithms, this yields
\begin{equation}
2\ln{|h|} - \ln{2} -\ln{\ln{|h|}} = \ln{a(|h|)} + \alpha(|h|)\ln{|h|}.\label{alpha_eq}
\end{equation} 
We now consider variations of $\alpha$ only, i.e. we fix $a$ with respect to $|h|$. 
Taking the derivative with respect to $|h|$ on both sides
of the equation and solving for $\alpha$ then yields
\begin{equation}
\alpha(|h|) \approx 2 - \frac{1}{\ln{|h|}}.
\end{equation} 
Inserting this result into Eq. (\ref{alpha_eq}), we finally obtain the estimate
\begin{equation}
a(|h|) \approx \frac{e}{2\ln{|h|}}
\end{equation} 
for the proportionality constant $a$.

\end{document}